\documentclass[12pt,preprint]{aastex62}

\usepackage{amssymb}
\usepackage{amsmath}
\usepackage{epstopdf}
\usepackage{color}
\usepackage{amsmath}
\usepackage{graphicx}
\usepackage{hyperref}
\usepackage{graphics,graphicx,natbib}

\def\lapp{\ifmmode\stackrel{<}{_{\sim}}\else$\stackrel{<}{_{\sim}}$\fi}
\def\gapp{\ifmmode\stackrel{>}{_{\sim}}\else$\stackrel{>}{_{\sim}}$\fi}

\newcommand{\presto}{{\tt{PRESTO}}}

\newcommand{\mb}{}
\newcommand{\vk}{}
\newcommand{\st}{}
\newcommand{\AJ}{}
\newcommand{\PC}{}
\newcommand{\zp}{}
\newcommand{\ef}{}

\begin{document}
\title{CHIME/FRB Discovery of Eight New Repeating Fast Radio Burst Sources} 
\shorttitle{}
\shortauthors{}

\author{The CHIME/FRB Collaboration}
\noaffiliation{}

\author{B.~C.~Andersen}
  \affiliation{Department of Physics, McGill University, 3600 rue University, Montr\'eal, QC H3A 2T8, Canada}
  \affiliation{McGill Space Institute, McGill University, 3550 rue University, Montr\'eal, QC H3A 2A7, Canada}
\author{K.~Bandura}
  \affiliation{CSEE, West Virginia University, Morgantown, WV 26505, USA}
  \affiliation{Center for Gravitational Waves and Cosmology, West Virginia University, Morgantown, WV 26505, USA}
\author{M.~Bhardwaj}
  \affiliation{Department of Physics, McGill University, 3600 rue University, Montr\'eal, QC H3A 2T8, Canada}
  \affiliation{McGill Space Institute, McGill University, 3550 rue University, Montr\'eal, QC H3A 2A7, Canada}
\author{P.~Boubel}
  \affiliation{Department of Physics, McGill University, 3600 rue University, Montr\'eal, QC H3A 2T8, Canada}
  \affiliation{McGill Space Institute, McGill University, 3550 rue University, Montr\'eal, QC H3A 2A7, Canada}
\author{M.~M.~Boyce}
  \affiliation{Department of Physics and Astronomy, University of Manitoba, Allen Building, Winnipeg, MB R3T 2N2, Canada}
\author{P.~J.~Boyle}
  \affiliation{Department of Physics, McGill University, 3600 rue University, Montr\'eal, QC H3A 2T8, Canada}
  \affiliation{McGill Space Institute, McGill University, 3550 rue University, Montr\'eal, QC H3A 2A7, Canada}
\author{C.~Brar}
  \affiliation{Department of Physics, McGill University, 3600 rue University, Montr\'eal, QC H3A 2T8, Canada}
  \affiliation{McGill Space Institute, McGill University, 3550 rue University, Montr\'eal, QC H3A 2A7, Canada}
\author{T.~Cassanelli}
  \affiliation{Department of Astronomy and Astrophysics, University of Toronto, 50 St.~George Street, Toronto, ON M5S 3H4, Canada}
  \affiliation{Dunlap Institute for Astronomy and Astrophysics, University of Toronto, 50 St.~George Street, Toronto, ON M5S 3H4, Canada}
\author{P.~Chawla}
  \affiliation{Department of Physics, McGill University, 3600 rue University, Montr\'eal, QC H3A 2T8, Canada}
  \affiliation{McGill Space Institute, McGill University, 3550 rue University, Montr\'eal, QC H3A 2A7, Canada}
\author{D.~Cubranic}
  \affiliation{Department of Physics and Astronomy, University of British Columbia, 6224 Agricultural Road, Vancouver, BC V6T 1Z1, Canada}
\author{M.~Deng}
  \affiliation{Department of Physics and Astronomy, University of British Columbia, 6224 Agricultural Road, Vancouver, BC V6T 1Z1, Canada}
\author{M.~Dobbs}
  \affiliation{Department of Physics, McGill University, 3600 rue University, Montr\'eal, QC H3A 2T8, Canada}
  \affiliation{McGill Space Institute, McGill University, 3550 rue University, Montr\'eal, QC H3A 2A7, Canada}
\author{M.~Fandino}
  \affiliation{Department of Physics and Astronomy, University of British Columbia, 6224 Agricultural Road, Vancouver, BC V6T 1Z1, Canada}
\author{E.~Fonseca}
  \affiliation{Department of Physics, McGill University, 3600 rue University, Montr\'eal, QC H3A 2T8, Canada}
\author{B.~M.~Gaensler}
  \affiliation{Dunlap Institute for Astronomy and Astrophysics, University of Toronto, 50 St.~George Street, Toronto, ON M5S 3H4, Canada}
  \affiliation{Department of Astronomy and Astrophysics, University of Toronto, 50 St.~George Street, Toronto, ON M5S 3H4, Canada}
\author{A.~J.~Gilbert}
  \affiliation{Department of Physics, McGill University, 3600 rue University, Montr\'eal, QC H3A 2T8, Canada}
  \affiliation{McGill Space Institute, McGill University, 3550 rue University, Montr\'eal, QC H3A 2A7, Canada}
\author{U.~Giri}
  \affiliation{Perimeter Institute for Theoretical Physics, 31 Caroline Street N, Waterloo, ON N2L 2Y5, Canada}
  \affiliation{Department of Physics and Astronomy, University of Waterloo, Waterloo, ON N2L 3G1, Canada}
\author{D.~C.~Good}
  \affiliation{Department of Physics and Astronomy, University of British Columbia, 6224 Agricultural Road, Vancouver, BC V6T 1Z1, Canada}
\author{M.~Halpern}
  \affiliation{Department of Physics and Astronomy, University of British Columbia, 6224 Agricultural Road, Vancouver, BC V6T 1Z1, Canada}
\author{A.~S.~Hill}
  \affiliation{Department of Physics and Astronomy, University of British Columbia, 6224 Agricultural Road, Vancouver, BC V6T 1Z1, Canada}
  \affiliation{Department of Computer Science, Math, Physics, and Statistics, University of British Columbia -- Okanagan, 3187 University Way, Kelowna, BC V1V 1V7, Canada}
  \affiliation{Space Science Institute, 4750 Walnut Street, Suite 205, Boulder, CO 80301, USA}
  \affiliation{Dominion Radio Astrophysical Observatory, Herzberg Astronomy \& Astrophysics Research Centre, National Reseach Council Canada, P.O.~Box 248, Penticton, BC V2A 6J9, Canada}
\author{G.~Hinshaw}
  \affiliation{Department of Physics and Astronomy, University of British Columbia, 6224 Agricultural Road, Vancouver, BC V6T 1Z1, Canada}
\author{C.~H\"ofer}
  \affiliation{Department of Physics and Astronomy, University of British Columbia, 6224 Agricultural Road, Vancouver, BC V6T 1Z1, Canada}
\author{A.~Josephy}
  \affiliation{Department of Physics, McGill University, 3600 rue University, Montr\'eal, QC H3A 2T8, Canada}
  \affiliation{McGill Space Institute, McGill University, 3550 rue University, Montr\'eal, QC H3A 2A7, Canada}
\author{V.~M.~Kaspi}
  \affiliation{Department of Physics, McGill University, 3600 rue University, Montr\'eal, QC H3A 2T8, Canada}
  \affiliation{McGill Space Institute, McGill University, 3550 rue University, Montr\'eal, QC H3A 2A7, Canada}
\author{R.~Kothes}
  \affiliation{Dominion Radio Astrophysical Observatory, Herzberg Astronomy \& Astrophysics Research Centre, National Reseach Council Canada, P.O.~Box 248, Penticton, BC V2A 6J9, Canada}
\author{T.~L.~Landecker}
  \affiliation{Dominion Radio Astrophysical Observatory, Herzberg Astronomy \& Astrophysics Research Centre, National Reseach Council Canada, P.O.~Box 248, Penticton, BC V2A 6J9, Canada}
\author{D.~A.~Lang}
  \affiliation{Perimeter Institute for Theoretical Physics, 31 Caroline Street N, Waterloo, ON N2L 2Y5, Canada}
  \affiliation{Department of Physics and Astronomy, University of Waterloo, Waterloo, ON N2L 3G1, Canada}
\author{D.~Z.~Li}
  \affiliation{Canadian Institute for Theoretical Astrophysics, University of Toronto, 60 St.~George Street, Toronto, ON M5S 3H8, Canada}
  \affiliation{Department of Physics, University of Toronto, 60 St.~George Street, Toronto, ON M5S 1A7, Canada}
\author{H.-H.~Lin}
  \affiliation{Canadian Institute for Theoretical Astrophysics, University of Toronto, 60 St.~George Street, Toronto, ON M5S 3H8, Canada}
\author{K.~W.~Masui}
  \affiliation{MIT Kavli Institute for Astrophysics and Space Research, Massachusetts Institute of Technology, 77 Massachusetts Ave, Cambridge, MA 02139, USA}
  \affiliation{Department of Physics, Massachusetts Institute of Technology, 77 Massachusetts Ave, Cambridge, MA 02139, USA}
\author{J.~Mena-Parra}
  \affiliation{MIT Kavli Institute for Astrophysics and Space Research, Massachusetts Institute of Technology, 77 Massachusetts Ave, Cambridge, MA 02139, USA}
\author{M.~Merryfield}
  \affiliation{Department of Physics, McGill University, 3600 rue University, Montr\'eal, QC H3A 2T8, Canada}
  \affiliation{McGill Space Institute, McGill University, 3550 rue University, Montr\'eal, QC H3A 2A7, Canada}
\author{R.~Mckinven}
  \affiliation{Dunlap Institute for Astronomy and Astrophysics, University of Toronto, 50 St.~George Street, Toronto, ON M5S 3H4, Canada}
  \affiliation{Department of Astronomy and Astrophysics, University of Toronto, 50 St.~George Street, Toronto, ON M5S 3H4, Canada}
\author{D.~Michilli}
  \affiliation{Department of Physics, McGill University, 3600 rue University, Montr\'eal, QC H3A 2T8, Canada}
  \affiliation{McGill Space Institute, McGill University, 3550 rue University, Montr\'eal, QC H3A 2A7, Canada}
\author{N.~Milutinovic}
  \affiliation{Department of Physics and Astronomy, University of British Columbia, 6224 Agricultural Road, Vancouver, BC V6T 1Z1, Canada}
  \affiliation{Dominion Radio Astrophysical Observatory, Herzberg Astronomy \& Astrophysics Research Centre, National Reseach Council Canada, P.O.~Box 248, Penticton, BC V2A 6J9, Canada}
\author{A.~Naidu}
  \affiliation{Department of Physics, McGill University, 3600 rue University, Montr\'eal, QC H3A 2T8, Canada}
  \affiliation{McGill Space Institute, McGill University, 3550 rue University, Montr\'eal, QC H3A 2A7, Canada}
\author{L.~B.~Newburgh}
  \affiliation{Department of Physics, Yale University, New Haven, CT 06520, USA}
\author{C.~Ng}
  \affiliation{Dunlap Institute for Astronomy and Astrophysics, University of Toronto, 50 St.~George Street, Toronto, ON M5S 3H4, Canada}
\author{C.~Patel}
  \affiliation{Department of Physics, McGill University, 3600 rue University, Montr\'eal, QC H3A 2T8, Canada}
  \affiliation{McGill Space Institute, McGill University, 3550 rue University, Montr\'eal, QC H3A 2A7, Canada}
\author{U.~Pen}
  \affiliation{Canadian Institute for Theoretical Astrophysics, University of Toronto, 60 St.~George Street, Toronto, ON M5S 3H8, Canada}
\author{T.~Pinsonneault-Marotte}
  \affiliation{Department of Physics and Astronomy, University of British Columbia, 6224 Agricultural Road, Vancouver, BC V6T 1Z1, Canada}
\author{Z.~Pleunis}
  \affiliation{Department of Physics, McGill University, 3600 rue University, Montr\'eal, QC H3A 2T8, Canada}
  \affiliation{McGill Space Institute, McGill University, 3550 rue University, Montr\'eal, QC H3A 2A7, Canada}
\author{M.~Rafiei-Ravandi}
  \affiliation{Perimeter Institute for Theoretical Physics, 31 Caroline Street N, Waterloo, ON N2L 2Y5, Canada}
\author{M.~Rahman}
  \affiliation{Dunlap Institute for Astronomy and Astrophysics, University of Toronto, 50 St.~George Street, Toronto, ON M5S 3H4, Canada}
\author{S.~M.~Ransom}
  \affiliation{National Radio Astronomy Observatory, 520 Edgemont Road, Charlottesville, VA 22903 USA}
\author{A.~Renard}
  \affiliation{Dunlap Institute for Astronomy and Astrophysics, University of Toronto, 50 St.~George Street, Toronto, ON M5S 3H4, Canada}
\author{P.~Scholz}
  \affiliation{Dominion Radio Astrophysical Observatory, Herzberg Astronomy \& Astrophysics Research Centre, National Reseach Council Canada, P.O.~Box 248, Penticton, BC V2A 6J9, Canada}
\author{S.~R.~Siegel}
  \affiliation{Department of Physics, McGill University, 3600 rue University, Montr\'eal, QC H3A 2T8, Canada}
  \affiliation{McGill Space Institute, McGill University, 3550 rue University, Montr\'eal, QC H3A 2A7, Canada}
\author{S.~Singh}
  \affiliation{Department of Physics, McGill University, 3600 rue University, Montr\'eal, QC H3A 2T8, Canada}
  \affiliation{McGill Space Institute, McGill University, 3550 rue University, Montr\'eal, QC H3A 2A7, Canada}
\author{K.~M.~Smith}
  \affiliation{Perimeter Institute for Theoretical Physics, 31 Caroline Street N, Waterloo, ON N2L 2Y5, Canada}
\author{I.~H.~Stairs}
  \affiliation{Department of Physics and Astronomy, University of British Columbia, 6224 Agricultural Road, Vancouver, BC V6T 1Z1, Canada}
\author{S.~P.~Tendulkar}
  \affiliation{Department of Physics, McGill University, 3600 rue University, Montr\'eal, QC H3A 2T8, Canada}
  \affiliation{McGill Space Institute, McGill University, 3550 rue University, Montr\'eal, QC H3A 2A7, Canada}
\author{I.~Tretyakov}
  \affiliation{Dunlap Institute for Astronomy and Astrophysics, University of Toronto, 50 St.~George Street, Toronto, ON M5S 3H4, Canada}
  \affiliation{Department of Physics, University of Toronto, 60 St.~George Street, Toronto, ON M5S 1A7, Canada}
\author{K.~Vanderlinde}
  \affiliation{Department of Astronomy and Astrophysics, University of Toronto, 50 St.~George Street, Toronto, ON M5S 3H4, Canada}
  \affiliation{Dunlap Institute for Astronomy and Astrophysics, University of Toronto, 50 St.~George Street, Toronto, ON M5S 3H4, Canada}
\author{P.~Yadav}
  \affiliation{Department of Physics and Astronomy, University of British Columbia, 6224 Agricultural Road, Vancouver, BC V6T 1Z1, Canada}
\author{A.~V.~Zwaniga}
  \affiliation{Department of Physics, McGill University, 3600 rue University, Montr\'eal, QC H3A 2T8, Canada}
  \affiliation{McGill Space Institute, McGill University, 3550 rue University, Montr\'eal, QC H3A 2A7, Canada}

\correspondingauthor{E. Fonseca}
\email{efonseca@physics.mcgill.ca}

\begin{abstract}
We report on the discovery of eight repeating fast radio burst (FRB) sources found using the Canadian Hydrogen Intensity Mapping Experiment (CHIME) telescope.  These sources span a dispersion measure (DM) range of 103.5 to 1281 pc cm$^{-3}$.  They display varying degrees of activity: six sources were detected twice, another three times, and one ten times. These eight repeating FRBs likely represent the {\ef bright and/or high-rate} end of a distribution of infrequently repeating sources. For all sources, we determine sky coordinates with uncertainties of $\sim$10$^\prime$. FRB 180916.J0158+65 has a burst-averaged DM = $349.2 \pm 0.3$ pc cm$^{-3}$ and a low DM excess over the modelled Galactic maximum (as low as $\sim$20 pc cm$^{-3}$); this source also has a Faraday rotation measure (RM) of $-114.6 \pm 0.6$ rad m$^{-2}$, much lower than the RM measured for FRB 121102. FRB 181030.J1054+73 has the lowest DM for a repeater, $103.5 \pm 0.3$ pc cm$^{-3}$, with a DM excess of $\sim$ 70 pc cm$^{-3}$. Both sources are interesting targets for multi-wavelength follow-up due to their apparent proximity. The DM distribution of our repeater sample is statistically indistinguishable from that of the first 12 CHIME/FRB sources that have not repeated. We find, with 4$\sigma$ significance, that repeater bursts are generally wider than those of CHIME/FRB bursts that have not repeated, suggesting different emission mechanisms. Our repeater events show complex morphologies that are reminiscent of the first two discovered repeating FRBs. The repetitive behavior of these sources will enable interferometric localizations and subsequent host galaxy identifications.
\end{abstract}


\section{Introduction}

Fast radio bursts (FRBs) are a transient astrophysical phenomenon consisting of millisecond-duration bursts of radio waves whose dispersion measures (DMs) imply cosmological origins \citep{lbm+07,tsb+13}. The physical mechanisms responsible for their emission are currently unknown, though many different models have been proposed \citep[see][]{pww+18,phl19}, ranging from synchrotron maser emission from young magnetars in supernova remnants \citep{lyu14,bel17,mms19} to cosmic string cusps \citep{bcv17}.

The discovery of the first repeating FRB source, FRB 121102, at a dispersion measure DM $\simeq$ 560~pc~cm$^{-3}$ \citep{sch+14,ssh+16a}, eliminated cataclysmic models as the only means for producing FRB emission. The repetitive nature of FRB 121102 enabled sub-arcsecond localization of the source via radio interferometry and subsequent optical identification of the low-metallicity host galaxy at redshift of 0.193 \citep{clw+17,tbc+17}, confirming its cosmological origin. Subsequent multi-wavelength imaging of the host galaxy resolved a star-forming region consistent with the position of FRB 121102, supporting the notion that FRBs are 
young, active compact objects \citep{bassa2017frb}. Moreover, high-radio-frequency observations of FRB 121102  demonstrated $\sim$100\% linearly-polarized bursts, with a large and declining Faraday rotation measure (RM) of $\sim$10$^5$ rad m$^2$ \citep{msh+18} that indicates an intensely magnetized environment. These properties, 
{\vk determined thanks to}
the repetitive nature and dedicated follow-up, have been used to formulate a model that posits a young magnetar embedded within a supernova remnant, itself residing within a low-mass, low-metallicity galaxy, as the source of FRB emission \citep{mbm17,mm18,mms19}.

FRB 121102 is also known for its complex pulse phenomenology \citep{ssh+16b,hss+18} which involves highly variable spectra. Until recently, all such observations had been carried out at radio frequencies above 700 MHz, with bursts detected as high as 8 GHz \citep{msh+18,gsp+18}. However, a recent burst from this source has been reported in the CHIME 400--800-MHz band \citep{jcf+19}, demonstrating low-frequency emission from this source for the first time.

As shown with FRB 121102, the discovery of additional repeating FRBs is important 
{\vk for several reasons.}
Radio-interferometric localizations enable a wide range of multi-wavelength studies of repeating-FRB host galaxies and any associated regions within positional uncertainties. Moreover, repeating FRBs enable radio follow up for determining polarization properties, as well as characterization of burst activity and repetetion rates \cite[e.g.,][]{oyp18}. A sample of repeating FRBs is also crucial for establishing similarities (or differences) in fundamental properties between repeating bursts and pulses from sources not yet observed to repeat. Any significant differences would reflect differences in underlying mechanisms that so far remain elusive due to the small number of known repeating sources. Finally, recent works have considered the possibility that most FRBs could originate from repeating sources \citep[e.g.,][]{csr+19,rav19}, and future detections will address this lingering question.

Recently, the CHIME/FRB Collaboration \citep{abb+19a,abb+19b} has reported the discovery of a second repeating FRB source, FRB 180814.J0422+73, at DM 189~pc~cm$^{-3}$, from which six bursts have been observed.  Interestingly, the source exhibits complex burst morphology and sub-burst downward frequency drifts strongly reminiscent of those seen in FRB 121102 \citep{hss+18} and FRB 170827 \citep{ffb18}.  This measurement established that FRB 121102 is not unique in its properties as a repeater, and that such behavior could be common to repeating FRB sources.  


In this work, we report the discovery of eight new repeating FRB sources with the CHIME telescope, ranging in DM from 103.5 to 1281~pc~cm$^{-3}$, and provide initial localizations (with precision $\sim$10$^\prime$). In Section \ref{sec:obs}, we discuss the observations taken with the CHIME telescope using the FRB-search instrument, as well as observations taken with a pulsar-timing backend in support of CHIME/FRB follow-up of these sources. In Section \ref{sec:analysis}, we describe the analysis of total-intensity and baseband data for the purposes of localization and burst characterization. In Section \ref{sec:discussion}, we present interpretations of our findings, both for per-burst properties and ensemble analyses. We summarize 
{\vk our} findings and conclusions 
in Section \ref{sec:conclusions}.
 
\section{Observations} 
\label{sec:obs}

\subsection{CHIME/FRB Detection of Repeating FRBs}
\label{sec:frbdetection}

A detailed description of the CHIME/FRB instrument was provided in an overview paper \citep{abb+18}.  Briefly, CHIME consists of four adjacent stationary cylinders of diameter 20~m and length 100~m, with axes oriented North-South to act as a transit telescope.  Each cylinder axis is populated with 256 dual-polarization antenna feeds, sensitive in the 400--800-MHz band, whose voltages are amplified, digitized and processed by an onsite FX-style correlator which feeds the FRB detection instrument. A realtime software pipeline identifies dispersed transient signals of millisecond durations in each of the 1024 formed sky beams, and buffered raw intensity data are dumped for all triggers having signal-to-noise ratio (S/N) greater than 10. Here we report on results during the interval from 2018 August 28 to 2019 March 13, when CHIME was in a commissioning state in which various components of the instrument were being tested, with software and calibration systems being frequently updated.   

We search for repeater candidates through our CHIME/FRB detection database (to be described elsewhere) by identifying bursts having sky positions that are coincident within one beamwidth (30$^\prime$ at 600 MHz) of each other, and having DMs within $\sim$10~pc~cm$^{-3}$ of each other.  The large DM window is chosen because the value measured by maximizing the pulse S/N is affected by the downward-drifting structure \citep{ssh+16b,hss+18} and because FRB pulses can sometimes have widths large enough to result in a fairly coarse DM measurement by our dedispersion code. Given the large amount of sky CHIME/FRB observes every day (the entire sky north of declination $-$10$^{\circ}$), as well as the range of DMs searched, we estimate the probability of two coincident events occurring by chance to be less than a few $\times 10^{-5}$ with the precise probability depending on exact assumptions (see Appendix~\ref{app:chancecoincidence} for details). Given the total number of events we have detected in the time range reported on here (to be described elsewhere), any two events coincident on the sky and in the stated DM range are unambiguously from the same source. We also require the events in question to have saved intensity data that permit us to examine the event dynamic spectrum, which, in the pipeline configuration used, has a S/N threshold of 10 for new sources, and 9 for events the pipeline suspects are repeat bursts. Refined DMs and positions (see below) are consistent within uncertainties.


During the aforementioned commissioning interval, we have detected eight new repeating sources, all listed in Table~\ref{ta:repeaters}.
For each source, in Appendix~\ref{app:exposure}, we show detailed 
plots of CHIME/FRB's exposure to its position, where the latter is highly dependent on declination.  Sources with
declination $>+70^{\circ}$ have much greater exposure, with circumpolar regions observed twice per day, in an upper
and lower transit.  However, note that the telescope's sensitivity in secondary transits is significantly reduced relative to that in the upper transit, and overall, sensitivity from day to day in our commissioning phase varied significantly, as discussed in \S\ref{sec:exposure}.


\begin{table}[t]
\begin{center}
\caption{Properties of Eight New CHIME/FRB Repeating Sources}
\centering
\resizebox{1.05\textwidth}{!}{ 
\hspace{-1.8in}
\begin{tabular}{cccccccccccc} \hline
    Source & Name$^a$  &  R.A.$^b$ & Dec.$^b$ & $l^c$ & $b^c$ & DM$^d$ & DM$_{\rm NE2001}^e$ & DM$_{\rm YMW16}^e$ & N$_{\rm bursts}$  & Exposure$^f$ & Completeness$\,^g$  \\
          &       &  (J2000)    &  (J2000) & (deg) & (deg) & (pc~cm$^{-3}$) & (pc~cm$^{-3}$) &(pc~cm$^{-3}$) &      &     (hr, upper / lower) &  (Jy ms)  \\\hline
1 &  180916.J0158+65 & 1h58m$\pm$7$'$   & +65$^\circ$44$'\pm$11$'$ & 129.7 &  3.7 & 349.2(3) & 200 & 330 & 10 & 23$\pm$8 & 4.2 \\
2 &  181030.J1054+73 & {\it 10h54m$\pm$8$'$}  & {\it +73$^\circ$44$'\pm$26$'$} & 133.4 & 40.9 & 103.5(3) & 40 & 32 & 2 & 27$\pm$14 / 19$\pm$11& ... / 17\\
3 &  181128.J0456+63 & {\it 4h56m$\pm$11$'$}  & {\it +63$^\circ$23$'\pm$12$'$} & 146.6 & 12.4 & 450.5(3) & 110 & 150 & 2 & 16$\pm$10 & 4.0\\
4 &  181119.J12+65 & 12h42m$\pm$3$'$  & +65$^\circ$08$'\pm$9$'$ & 124.5  & 52.0 & 364.05(9) & 34 & 26 & 3 & 19$\pm$9& 2.6 \\
  &                        & 12h30m$\pm$6$'$ &
+65$^\circ$06$'\pm$12$'$ &        &      & &    &    &   &   \\
5 &  190116.J1249+27 & 12h49m$\pm$8$'$  & +27$^\circ$09$'\pm$14$'$ & 210.5 & 89.5 & 441(2) & 20 & 20 & 2 & 8$\pm$5& 5.7\\
6 &  181017.J1705+68 & {\it 17h05m$\pm$12$'$} & {\it +68$^\circ$17$'\pm$12$'$} & 99.2 & 34.8 & 1281.6(4) & 43 & 37 & 2 & 20$\pm$11 & 5.6\\
7 & 190209.J0937+77 & {\it 9h37m$\pm$8$'$}   & {\it +77$^\circ$40$'\pm$16$'$} & 134.2 & 34.8 & 425.0(3) & 46 & 39 & 2 & 34$\pm$19 / 28$\pm$18  & 3.8 / ... \\
8 & 190222.J2052+69 & 20h52m$\pm$10$'$ & +69$^\circ$50$'\pm$11$'$ & 104.9 & 15.9 & 460.6(2) & 87 & 100 & 2 & 20$\pm$10&5.4 \\\hline
\end{tabular}
}
\label{ta:repeaters}
\end{center}
$^a$ Here we employ the naming convention (YYMMDD.JHHMM$\pm$DD) used in \cite{abb+19a} and \cite{abb+19b} in the current absence of a final naming convention agreed upon by the community.  These names therefore are likely to change.  The date in the name corresponds to our first detection of the source.  For brevity, and for the remainder of the paper, we refer to the repeaters by Source number (Column 1). For sources with non-contiguous error regions, the name is defined by the central position, except for Source 4, for which the `central' R.A. is not well defined at the minute level (see Figure~\ref{fig:localization}).\\
$^b$ Positions were determined from intensity data, except for Sources 1, 5 and 8, which were also informed from an analysis of baseband data (see \S\ref{sec:localization}).  Sources with position in italics have three or more non-contiguous error regions, with the tabulated position the central region only, with 90\% confidence uncertainty regions.  See Figure~\ref{fig:localization} for details. Source 4 has two non-contiguous uncertainty regions, resulting in two position entries (see Fig.~\ref{fig:localization}).\\
$^c$ Galactic longitude and latitude for the best position.\\
$^d$ Weighted average DM (see Table~\ref{ta:bursts}).\\
$^e$ Maximum model prediction along this line-of-sight
for the NE2001 \citep{ne2001} and YMW16 \citep{ymw17} Galactic electron density distribution models. Neither model accounts for DM contributions from the Galactic halo, which can be up to 50--80 pc cm$^{-3}$ \citep{prochaska2019probing}.\\
$^f$ For sources observed twice a day, the second entry corresponds to the less sensitive lower transit.
The uncertainties in the total exposure for the upper and lower transits of each source are dominated by the corresponding source declination uncertainties since the widths of the stationary synthesized beams vary significantly with declination
 (see \S\ref{sec:exposure}).\\
$^g$ Fluence completeness limits are given at the 90\% confidence level (see~\S\ref{sec:completeness}). For sources with two transits, we compute a completeness only where bursts were observed. The difference in sensitivity between upper and lower transits is roughly a factor of four for Source 2 and a factor of three for Source 7.
\end{table}
\normalsize

\subsection{CHIME/FRB Baseband Data}
\label{sec:baseband}

As described in \citet{abb+18}, 
the CHIME/FRB pipeline includes 
a system for saving buffered, channelized baseband voltage data upon a trigger by a bright FRB from the realtime search engine.
To trigger a dump of the baseband buffer, we currently require the
source to have a DM consistent with being extragalactic, S/N$>$15, not flagged as RFI, and
having a tree index \citep[see][]{abb+18} of $\leq 2$.
When these conditions are met, 
baseband data around the pulse are stored for each of the
1024 spectral frequencies and all 2048 digital correlator inputs.
Downsampling of trial DMs in our dedispersion code {\tt bonsai} \citep{abb+18} leads to a significant DM uncertainty
for the initial trigger, which in turn induces a timing uncertainty at
frequencies away from 400\,MHz (where trigger times are referenced).  While at 400 MHz we store 100 ms of baseband data, to account for this uncertainty, we
store additional data at higher frequencies, approaching a maximum of  $\sim$300\,ms 
at 800\,MHz.

\subsection{CHIME/Pulsar Instrument}
\label{sec:pulsar}

As mentioned in \citet{abb+18} and \citet{n++17}, the CHIME telescope has also been outfitted with a GPU-based backend that accepts data streams for 10 dual-polarization, tied-array beams from the CHIME X-Engine. After an FRB repeater candidate is discovered by CHIME/FRB, the CHIME/Pulsar backend monitors the nominal coordinates whenever possible, tracking the source position as it drifts through the primary beam of CHIME.  In total, 278.6 hours were spent on all repeaters reported in this paper using the CHIME/Pulsar backend, with the breakdown of time detailed in \S\ref{sec:pulsardetection}. The CHIME/Pulsar backend coherently dedisperses the beamformed data at the nominal DM found in the initial discovery. These data are recorded at a time resolution of 327.68~$\mu$s with 1024 frequency channels each with a width of 390\,kHz. The two polarizations are summed and analyzed offline with a \textsc{presto}-based single pulse search algorithm\footnote{https://github.com/scottransom/presto}.

\section{Analysis \& Results}
\label{sec:analysis}


\subsection{Source Localization}
\label{sec:localization}

\begin{figure}[b]
	\centering
\includegraphics[scale=0.45]{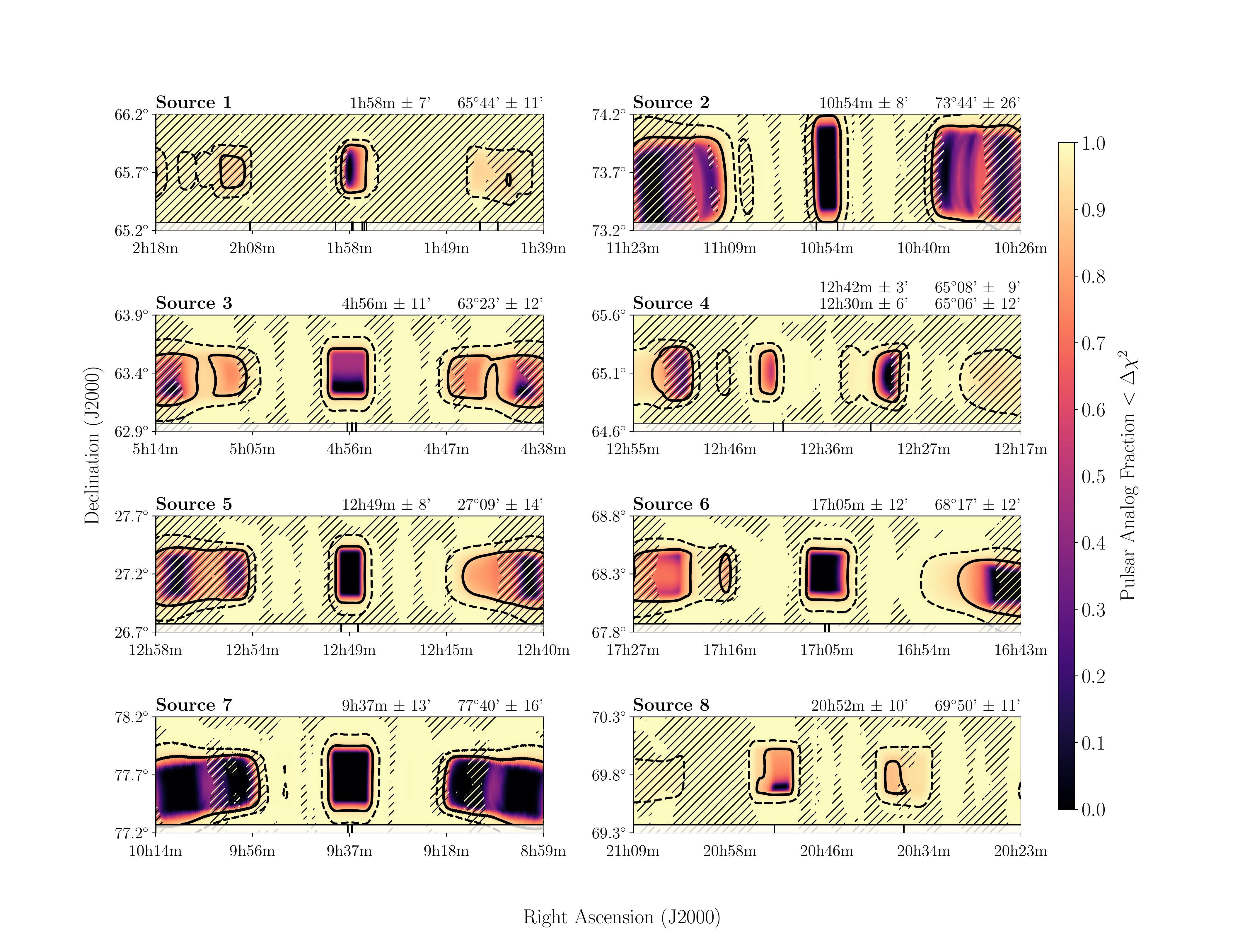}
\figcaption{Detection positions of the new CHIME/FRB repeating FRB sources, as determined from CHIME/FRB detection beam information through the methods described in \S\ref{sec:localization}. Each panel is $1^\circ \times 4^\circ$. Localization is performed as a $\chi^2$-minimization. The method is applied to a large population of analogous pulsar events (i.e., pulsars with similar brightness and beam-detection statistics), which we use to translate $\Delta\chi^2$ values to empirical confidence intervals depicted by the color scale. The 90\% and 99\% confidence intervals are indicated as solid and dashed contours; we use the former interval to report the most likely positions. The R.A. of the beam centers for each detection are shown as black ticks on the bottom of each panel. Hatched regions represent areas where, for at least one burst in the sample, the beam model predicts low sensitivity in the portion of the band where emission is observed --- see text for details.}
\label{fig:localization}
\end{figure}

Here we describe the methods used
to determine sky positions for these newly discovered
repeaters, combining information from
the multiple detections per source.
Our best estimates of our source positions are provided in Table~\ref{ta:repeaters}, and graphical depictions of the localization regions are
shown in Figure~\ref{fig:localization}.

As described in \citet{abb+19b}, 
we compare per-beam S/N with beam-model predictions, allowing localization to be carried out as a grid search $\chi^2$-minimization, with confidence intervals that are expressed with contours of constant $\Delta\chi^2$. While the underlying methodology is the same as in our previous work, we have made two important changes to the process. First, we have updated our beam-model to include East-West aliasing of the synthesized beams, as well as an approximated forward-gain model of the primary beam, which is based on ray-tracing simulations of the CHIME Pathfinder \citep{baa+14}. Second, we are now using two S/N values per-beam, which correspond to the trial spectral indices searched by the dedispersion engine (L1) \citep[see][]{abb+18}. This feature was not active in the early stages of pre-commissioning. For both changes, localization tests with pulsars show improvements that, while significant, are largely confined to sidelobe detections. This is expected, since these changes are most relevant for lines of sight with strong chromatic attenuation. {\AJ Further improvements to the beam-model are under active development. In particular, holography observations of both steady sources and pulsars are being used to refine the model of the primary beam \citep{bna+16}.}

To estimate systematic uncertainties, we apply our localization method to a large sample of pulsar events.  We start with all events on 2019 March 1 that have been associated with known pulsars by the real-time pipeline. {\AJ Note that preliminary studies of transiting calibration sources have shown the pointing of the synthesized beams to not exhibit time-variability.} {\AJ We remove PSR B0329+54 due to non-linearities between fluence and detection S/N observed for sources with extreme brightness}, as well as multi-beam events that include more than two beams (to match the repeater bursts we report), and events for which the true position is more than 2.5$^{\circ}$ from the meridian. The resulting sample includes ${\sim}$20k single-beam and ${\sim}$10k two-beam events, which are distributed among 193 pulsars that cover a declination range of $-$10$^\circ$ to +70$^\circ$. No strong declination-dependent effects are apparent in terms of localization accuracy.

We record the $\Delta\chi^2$ for each event at the true position within the grid of trial positions. If our statistical uncertainties dominate, we expect 99\% of the test sample to have a $\Delta\chi^2$ value less than the threshold used for the 99\% confidence interval. Instead, we find that roughly 80\% of the 30k events are contained, 
suggesting there are systematic sources of uncertainty that have not been accounted for
in our statistical treatment.  In order to measure the uncertainty directly,
we use percentiles of the test distribution to set $\Delta\chi^2$ thresholds. On the other hand, roughly 88\% of single-beam events are contained in the 90\% interval, suggesting that the empirical thresholds should be set according to specific detection scenarios.
We note that this empirical strategy for measuring uncertainties takes into account known
and unknown sources of error that are present for broadband sources such as the pulsars that are used for measurement. Narrow-band FRBs may have other sources of localization
error that are unaccounted for herein.

To simulate each repeater case, we draw a large number of analogously composed sub-samples from the test sets (e.g., one two-beam and two single-beam detections for Source~3).  These are drawn randomly (and with replacement).  We do not require that the events come from the same pulsar. The $\Delta\chi^2$ values are summed for each sub-sample and we use percentiles of the resulting underlying source location distribution to get empirical confidence intervals appropriate for each repeater. In each case, we use these intervals for quoted uncertainties. 

As we have used only band-averaged S/N values for a small number of events in determining our localizations, unsurprisingly, degeneracies persist in many cases.  Localization based on raw intensity data will improve on this, but this approach is still under development. 
In the interim, we can place some additional constraints based on the extent of the observed emission for each burst. For example, if a burst is visually confined to 400--500 MHz, we rule out locations for which the beam model predicts the sensitivity in the relevant range to be heavily attenuated (relative to the rest of the band).  
We require the median sensitivity at frequencies where emission is
clearly observed to be more than 5\% of the peak sensitivity across the
full 400-MHz band, leading to a mask of allowed locations for
individual bursts.
These additional constraints are most relevant when considering the possibility of detecting narrow-band bursts within the sidelobes, where the sensitivity is highly chromatic (East-West alias position at 400~MHz is ${\sim}2^\circ$ away from the synthesized beam center, whereas the offset at 800~MHz is ${\sim}1^\circ$). 

The available baseband data (see Table~\ref{ta:bursts}) allowed us to confirm the localization for Sources 1, 5 and 8.
We produced a grid of closely-spaced tied-array beams around the best position and selected the one where the S/N was maximum.
In addition, multiple locations had comparable probabilities for Source 5 and 8 (see Figure~\ref{fig:localization}). 
Therefore, a grid of tied-array beams contiguous at their FWHM were produced to cover the whole uncertainty regions.
In all cases, the baseband analysis confirmed the position having the highest probability in Figure~\ref{fig:localization}.
A refined localization based on fitting baseband data with a beam model is in progress and will be presented elsewhere.


\subsection{Exposure Determination}
\label{sec:exposure}

In order to estimate the exposure of the CHIME/FRB system to the sources reported in this work, we add up the duration of the daily transits of each source across the FWHM region of the synthesized beams at 600 MHz. We include transits in the interval from 2018 August 28 to 2019 February 25 when the CHIME/FRB system was in a commissioning phase. The transits for the pre-commissioning phase (2018 July 25 -- 2018 August 27) were not included as the difference in the synthesized beam configuration resulted in the sensitivity to a given sky location being significantly different between the two phases. Additionally, we do not include transits observed after 2019 February 25 although we are reporting on bursts detected since then. This is because major upgrades were being made to the software pipeline through 2019 March resulting in large sensitivity variations which cannot be adequately characterized.

For each source, sky positions within the uncertainty regions shown in Figure \ref{fig:localization} have different transit times across the synthesized beams due to their different declinations and hence elevation angles. 
Therefore, we generate a uniform grid of locations within the  90\% confidence uncertainty region and for each trial position, calculate the total duration for the transits during which the CHIME/FRB pipeline was fully operational. 
We exclude intervals during which observations were interrupted by commissioning activities or issues with computing nodes processing data for the beams through which the source transits. To obtain the total exposure, we calculate the weighted average over all trial positions with the weights equal to the percentage probability of the source being located at a given position.  For each source, the weighted mean and standard deviation of the exposure are reported in Table \ref{ta:repeaters} and shown graphically in the Appendix (Figure~\ref{fig:exposure}). High-declination sources ($\delta > +70^\circ$) have the exposure for the lower transit reported separately. The large uncertainties in the exposures are due to the CHIME/FRB system having a tiling of synthesized beams that are not touching at the half-power points at 600 MHz. The exposure above the half-power level to any astrophysical source transiting across a synthesized beam is then highly dependent on the declination of the source as it cuts through that two-dimensional beam. The exposure also depends significantly on observing frequency, due to the FWHM regions of the synthesized beams being larger at low frequencies.


\subsection{Burst Fluence \ef{and Peak Flux} Determination}
\label{sec:fluence}


We used the calibration methods described in \citet{abb+19a}, \citet{abb+19b},
and \citet{jcf+19} to determine fluences for all bursts
reported on here. {\vk }Briefly, we used several bright point sources in the vicinity of each burst to calibrate, and their variations to ascertain the calibration error. The only differences were:
\begin{itemize}
    \item For Sources 1, 3, 4 and 5, we used calibration sources within 5$^{\circ}$ of declination from the FRB. For calculating the uncertainty, we formed pairs of sources by choosing sources within 5$^{\circ}$ in declination from each other and estimated the fluence uncertainties as we did in previous work.
    \item For Sources 6, 7 and 8, we used calibration sources in a 5$^{\circ}$ declination  range on the opposite side of the zenith, since there are no calibration sources within 5$^{\circ}$ of these sources. For calculating the uncertainties, we chose sources on opposite sides of the zenith and estimated the fluence uncertainty again as in previous work. 
    \item For Source 2, we used calibration sources in a 5$^{\circ}$ declination angle range on the opposite side of the zenith but then multiplied the resultant fluence by a factor of four, since the bursts were detected in the lower transit \citep[see][]{abb+19b}. 
\end{itemize}
{\ef The peak fluxes for each burst were determined using the same calibration scheme as the corresponding fluence calculation. The peak flux was taken to be the maximal flux value within the extent of the burst (binned at 0.98304 ms) in the band-averaged time series, with an uncertainty derived from our calibration sources as described above. If there were multiple components in a given burst, then a peak flux measurement was obtained for each sub-burst.}

\subsection{Fluence Completeness Determination}
\label{sec:completeness}

The sensitivity of the CHIME/FRB system to a burst from a particular sky location varies
during the observing time reported in Table~\ref{ta:repeaters}. There are three potential sources of this variability, namely the day-to-day variations due to changes in gain-calibration strategies and the software pipeline, varying response of synthesized beams over the duration of the source transit and their complex bandpass
resulting in effective sensitivity that is strongly dependent on the emission frequencies and bandwidths of individual bursts. Additionally, for high declination sources ($\delta > +70^\circ$), the sensitivity varies significantly between the upper and lower transit due to reduced primary beam response at lower elevation angles. We estimate this reduction to be a factor of $\sim$4 at the declination of Source 2 \citep[see][]{abb+19b}. For Source 7, the reduction is slightly lower, a factor of $\sim$3. 

To compute a fluence completeness limit across a quoted exposure, we simulate a
large ensemble of fluence thresholds for different detection scenarios,
following the methods detailed in \citet{jcf+19}.
Each realization includes an epoch, position along transit, and a Gaussian spectral model,
which are used to estimate the relative sensitivity between the simulated and real detections.
The relative sensitivity 
is then used to scale the fluence threshold inferred from the real detection to get a simulated threshold.
To get a fluence completeness limit at the 90\% confidence level, we take the 90th percentile of the distribution of simulated fluence thresholds, such that 90\% of simulated events would be detectable above the corresponding fluence.
To characterize day-to-day variation, we analyze S/N values for pulsars
that have a declination within 5$^{\circ}$ of the source and are reliably detected by
CHIME/FRB. Typical daily variations are at the 20\% level.
Intra-transit variation is characterized with band-averaged sensitivity
predictions from a beam model. Because of the FWHM definition of our exposure (see~\S\ref{sec:exposure}),
transit sensitivity may vary by a factor of two.
We characterize intra-band variation with the beam-former-to-Jansky calibration products
used for fluence determination \citep[see][]{abb+19a}. Depending on the declination of
the source and the emission bandwidths considered, these variations may
span a factor of two as well.
%
%
%
To extend the methodology, originally developed for a single burst from FRB~121102, we have made the following changes:

\begin{itemize}
    \item To include multiple bursts, we associate each realization with a randomly
            selected burst from the source, which is used as the reference for determining
                the relative sensitivity and initial fluence threshold.
    \item To estimate the daily sensitivity variation for sources with declination
            $> +70^\circ$, we select pulsars
                within 10$^{\circ}$ of the source declination,
                rather than 5$^{\circ}$, to increase otherwise sparse sample
                sizes. Additionally, for the lower transit, we assumed beam
                symmetry in the North-South direction and included pulsars
                detected on the opposite side of the zenith.
    \item To handle uncertainties in source location, we associate each
            realization with sky coordinates randomly drawn according to the
                probability distribution shown in Figure~\ref{fig:localization}. These
                coordinates are used to define the transit path and
                scale initial fluence measurements up, as they are originally referenced
                to the center of the synthesized beam.
\end{itemize}

\subsection{Burst Properties}
\label{sec:bursts}


We used a variety of techniques to characterize burst properties.
For all bursts, structure-optimizing DMs {\zp \citep[i.e., the DM for which all significant sub-burst emission arrives simultaneously rather than the DM that maximizes S/N, see also][]{hss+18}} were fit by maximizing the coherent power in the pulse across the emission bandwidth using the \texttt{DM\_phase} package\footnote{\url{https://github.com/danielemichilli/DM_phase}} (Seymour et al. in prep.). The algorithm calculates a ``coherence spectrum'' for a range of trial DM values by taking a one-dimensional Fourier transform of the intensity data along the frequency channels, dividing by the amplitude (thus keeping only the phase 
information) and summing over the emission bandwidth. Sub-bursts that line up (i.e., are coherent) will have similar phase at a given fluctuation frequency and will sum more coherently to a greater amplitude. To optimize the sharpness of the sub-bursts, we are interested in the {time-derivative of the burst profile, so the power spectrum is calculated by summing over the fluctuation frequencies, multiplied by the squares of those frequencies. 
An example of coherence spectra, multiplied by the squared frequencies, for 100 trial DM values is shown on the bottom left in Figure~\ref{fig:dmphase}, 
{\zp showing only the} range of fluctuation frequencies away from short- and long-{\zp fluctuation-}frequency noise. The final power spectrum is shown in the top left of the figure. The DM at which a {\zp high-order} polynomial fit to the power spectrum peaks is the structure-optimizing DM. The statistical DM uncertainty is calculated from the probability density function associated with the power spectrum, assuming a uniform distribution in phase angles. We have verified the method by using it to determine the DMs of bursts from various known pulsars, for both CHIME/FRB intensity and baseband data. 

The alignment of sub-bursts in the dynamic spectra, dedispersed to the DM found by this method, were verified by eye (see, e.g., Figure~\ref{fig:dmphase}) and for bursts with sufficiently high S/N, we verified that a forward-derivative structure-optimizing method \citep{gsp+18,hss+18,jcf+19} results in a similar DM measurement. For the 181222 burst of Source 1 -- which has high S/N and exhibits sharp features -- we measure the most precise DM, with an uncertainty of 0.1 pc cm$^{-3}$ (calculated from the probability density function associated with the power spectrum). This uncertainty corresponds to a dispersion delay of 0.5 ms from 800 to 600 MHz (the emission bandwidth of this burst), half the sampling time of the data. In case a burst's S/N was too low for \texttt{DM\_phase} to converge, we resorted to measuring an S/N-optimizing DM instead. For a sharp single-component burst and in the limit of low S/N, the structure-optimizing and S/N-optimizing algorithms converge to the same DM values.

Dynamic spectra (``waterfall'' plots) for all bursts are presented in Figure~\ref{fig:waterfall} for the best estimate of each burst's DM, as provided in Table~\ref{ta:bursts}.  For none of our repeating sources, including Source 1 for which we observed 10 bursts over 4 months, do we detect any significant change in DM.

We attempt to measure 
the linear drift rate of
downward-drifting sub-bursts, as observed for FRB 121102 and FRB 180814.J0422+73, using an auto-correlation analysis \citep{hss+18}. Integrated auto-correlation results are fit with Gaussian profiles using a least-squares optimization routine
described by \citet{jcf+19}. To improve the robustness of the auto-correlation analysis, we have added a Monte Carlo resampling step: we draw 100 random DM values from the DM uncertainty distribution, dedisperse the burst to that DM, and refit the linear drift rate 100 times, after having added a random instance of noise to our data model.
Noise is added in the auto-correlation space (as this is less computationally intensive than adding noise to the burst model in frequency-time and calculating the 2D auto-correlation for each iteration). The initial tilted 2D Gaussian fit to the 2D auto-correlation results is our data model and noise is drawn according to the measured statistical properties of the 2D auto-correlation of an off-burst region. We verified that the residuals, after subtracting the model from the data, were distributed as the noise from the off-burst region. Only for the 181222 burst from Source 1 was this not the case, as the high S/N sharp components in this burst make a tilted 2D Gaussian an insufficient model for describing the 2D auto-correlation of the burst. For this burst we instead resampled the data in time-frequency space, where we take the multi-component model for the burst (see below) and add noise drawn from the distribution of an off-burst region to the model before dedispersing the data to a random DM value from the DM uncertainty distribution and proceeding with the 2D auto-correlation.
Results for the drift rates, when measurable, are presented in Table \ref{ta:bursts}; the quoted uncertainty is the  interval that contains 68\% of the linear drift rate measurements across all Monte Carlo realizations. Note that the marginalized drift rate distributions are non-Gaussian due to the covariance between DM and drift and that the 68\% containment regions thus do not fully describe the underlying distribution.
As can be seen from Table~\ref{ta:bursts}, we measure a significant drift in eleven cases: for events 181019, 181222, 181223, 181226 (Source 1), 181128, 181219 (Source 3), 190103 (Source 4), 190116A (Source 5), 190216 (Source 6), 190210 (Source 7) and 190301 (Source 8).

All dynamic spectra shown in Figure~\ref{fig:waterfall} were also analyzed with a separate burst-fitting algorithm used by \citet{abb+19a} and \citet{jcf+19}. We specifically used the algorithm implementation presented by Josephy~et.~al. that allows for an arbitrary number of two-dimensional profile components to be fit against a given spectrum, along with the fitting of ``global" parameters such as the DM and scattering timescale. While all arrival times and temporal widths are fitted with Gaussian profiles, we fitted Gaussian or ``spectral" profiles (i.e., a power-law distribution with a running spectral index) along the frequency axis. For each burst in Table~\ref{ta:bursts}, the number of profile components ($n$) was determined by comparing improvements in the goodness of fit (i.e., $\Delta\chi^2$) between best-fit multi-component models; the number $n$ was determined by noting the first instance where $(\Delta\chi^2_{n+1}-\Delta\chi^2_n) < 5$, which is the minimum number of parameters needed to model a two-dimensional burst component. In order to ensure adequate and uniform modeling of the repeater bursts -- especially those with sub-bursts -- we held the DM fixed to the values determined by either the coherence-spectrum or S/N-maximizing dedispersion methods discussed above. The values of scattering timescales and widths shown in Table~\ref{ta:bursts} were determined using this burst-fitting procedure.

As can be seen in Table \ref{ta:bursts}, most scattering times are reported as upper limits.  For Source 1, we note that event 181223's measured scattering time appears to be inconsistent with the limit set by event 181226. However, we caution that intrinsically complex morphology -- which varies between bursts -- can bias scattering-timescale measurements towards high values. Therefore the apparent inconsistency in scattering estimates between different bursts from Source 3 does not necessarily signal variability of scattering properties in this source.


\begin{figure}[t]
	\centering
\includegraphics[width=0.8\textwidth]{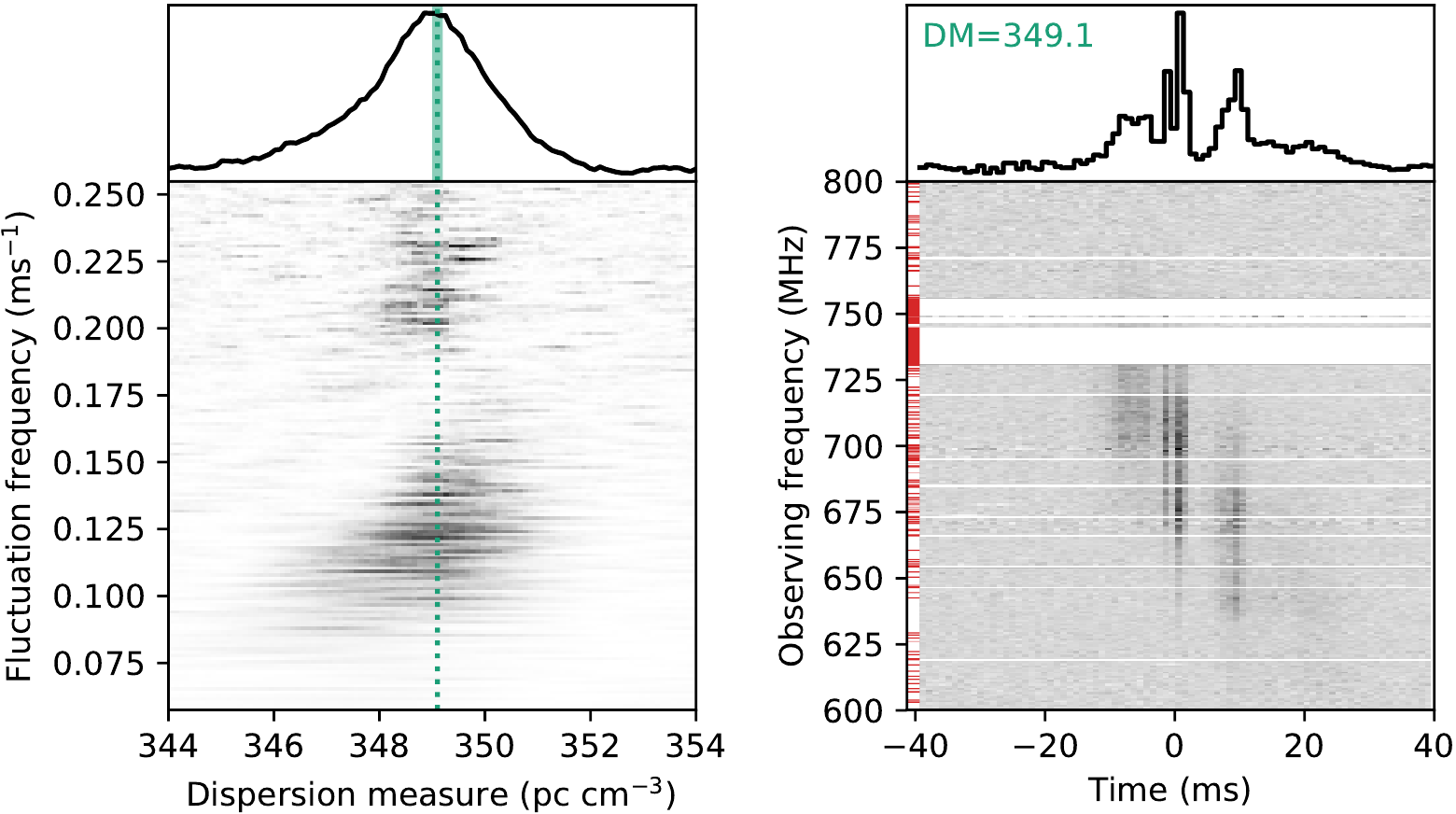}
\figcaption{Left: Coherent power in fluctuation frequency based on the full-resolution data in the 600--800 MHz range multiplied by fluctuation frequency squared as a function of trial DM value for the 181222 burst from Source 1, with the power spectrum (the sum of the coherent power weighted by frequency squared) on the top. The optimal DM and statistical uncertainty (calculated from the probability density function associated with the power spectrum) are indicated with a green dotted line and shaded region. Right: {\zp 0.98304-ms time resolution b}urst intensity data dedispersed to the DM {\zp (in pc cm$^{-3}$)} that maximizes the coherent power in the pulse and downsampled in frequency by a factor 32 {\zp to a {\vk frequency resolution of} 781.25 kHz}.
The sharp alignment of sub-bursts is visible. Underlying missing or masked channels in the full-resolution intensity data are represented by red lines on the left of the intensity data.}
\label{fig:dmphase}
\end{figure}

\begin{table}[t]
\begin{center}
\caption{{\zp Individual Burst} Properties from the Eight New CHIME/FRB Repeaters.$^a$ }
\hspace{-1.in}
\resizebox{1.1\textwidth}{!}{ 
\begin{tabular}{cccccccccc} \hline
    Day & MJD & Arrival Time$^b$  &  DM & Drift Rate & Width$^b$ & Scattering Time & Fluence & \ef{Peak Flux} \\
  (yymmdd) &     & (UTC @ {\ef 600} MHz)  & (pc~cm$^{-3}$) & (MHz/ms) & (ms) & (ms @ 600 MHz)& (Jy ms) & \ef{(Jy)} \\\hline
        \multicolumn{7}{c}{Source 1} \\\hline
 180916 & 58377 & 10:15:19.8021(2) & 349.2(4) & ... & 1.40(7) & $<1.5$ & 2.3(1.2) & 1.4(6) \\
 181019 & 58410 & 08:13:22.7507(8) & 349.0(6) &  $-28_{-26}^{+9}$ & 4.1(3) / 4.4(9) & $<4.7$ & 2.7(1.3) & 0.6(3) / 0.3(2)\\ 
 181104A & 58426 & 06:57:18.58524(12) & 349.5(3) & ... & 1.37(7) & $<1.5$ & 2.5(1.2) & 1.4(5) \\ 
 181104B & 58426 & 07:07:01.591(4) & 349.6(2)$^d$ & ...  & 6.3(1.1) & $<3.5$ & 2.0(1.0) & 0.4(2) \\
 181120 & 58442 & 05:56:06.23243(9) & 349.9(6) & ... & 1.10(9) & $<1.3$ & 1.8(8) & 1.1(5) \\
 181222 & 58474 & 03:59:23.2082(3) & 349.1(1) & $-4.6_{-0.2}^{+0.2}$ & 4.95(4) / 1.51(3)  & $<1.6$ & 27(12) & 1.7(7) / 4.9(1.8) \\
 & & & & & 3.7(3) / 2.8(3) & & & 3.0(1.0) / 0.7(3) \\
 181223 & 58475 & 03:51:28.96040(17) & 349.7(7) & $-30_{-34}^{+11}$ & 1.06(5) / 6.3(5) & 2.4(3) & 8.1(3.8) & 1.7(6) / 0.5(3) \\ 
 181225$^c$ & 58477 & 03:53:03.9260(4) & 348.9(7) &  ... & 1.3(3) & 2.0(5) & 1.9(9) & 0.4(2) \\
 181226$^c$ & 58478 & 03:43:30.1074(2) & 348.8(2) & $-20_{-16}^{+4}$ & 0.87(3) / 3.6(4) & $<0.9$ & 3.8(1.8) & 1.6(6) / 0.6(3) \\
 190126 & 58509 & 01:32:45.3289(3) & 349.8(5) & ... & 2.53(13) & $<2.8$ & 2.0(1.0) & 0.7(3) \\\hline
 \multicolumn{7}{c}{Source 2} 
 \\\hline
 181030B & 58421 & 04:13:13.0255(6) & 103.5(7) & ... & 0.59(8) & $<0.8$ & 7.3(3.8) & 3.2(1.7) \\
 181030B & 58421 & 04:16:21.6546(14) & 103.5(3) & ... & 1.43(8) & $<1.6$ & 4.5(1.8) & 3.1(1.4)
 \\\hline
 \multicolumn{7}{c}{Source 3} \\\hline
 181128 & 58450 & 08:27:41.7400(5) & 450.2(3) & $-11_{-2}^{+21}$ & 2.43(16) / 5.4(6) & $<2.8$ & 4.4(2.2) & 0.5(3) / 0.3(2) \\
 181219 & 58471 & 07:04:41.6780(9) & 450.8(3) & $-14_{-6}^{+3}$ & 5.5(7) & $<6.9$ & 2.5(1.2) & 0.3(2)
 \\\hline
 \multicolumn{7}{c}{Source 4} \\\hline
 181119 & 58441 & 16:49:03.1914(8) & 364.2(1.0) & ... & 6.3(6) & $<7.5$ & 1.8(0.8) & 0.3(2)\\
 190103 & 58486 & 13:47:23.3225(5) & 364.0(3) & $-22_{-4}^{+3}$ & 2.66(10) & $<2.8$ & 2.5(1.2) & 0.6(3) \\
 190313 & 58555 & 09:21:46.7250(6) & 364.2(6) & ... & 1.5(2) & $<1.9$ & 1.0(5) & 0.4(2)\\\hline
 \multicolumn{7}{c}{Source 5} \\\hline
 190116A$^c$ & 58499 & 13:07:33.833(1) & 444.0(6) & $-14_{-4}^{+4}$ & 4.0(5) & $<11$ & 0.8(4) & 0.3(2) \\
 190116B & 58499 & 13:08:20.4129(2) & 443.6(8) & ... & 1.5(3) & $<1.7$ & 2.8(1.4) & 0.4(2)
 \\\hline
   \multicolumn{7}{c}{Source 6} \\\hline
 181017 & 58408 & 23:26:11.8600(16) & 1281.9(4) & ... & 13.4(1.4) & $<16$ & 1.0(5) & 0.4(3) \\
 190216 & 58530 & 15:26:58.029(2) & 1281.0(6)$^d$ & $-1.9_{-0.3}^{+0.2}$ & 20.2(1.7) & 11(2) & 16(5) & 0.4(2) 
 \\\hline
   \multicolumn{7}{c}{Source 7} \\\hline
 190209 & 58523 & 08:20:20.977(1) & 424.6(6) & ... & 3.7(5) & $<4.7$ & 2.0(1.0) & 0.4(2) \\
 190210 & 58524 & 08:17:13.907(3) & 425.2(5) & $-12_{-7}^{+4}$ & 9.4(1.4) & $<12$ & 0.5(3) & 0.6(4) 
 \\\hline
   \multicolumn{7}{c}{Source 8} \\\hline
 190222 & 58536 & 18:46:01.3679(2) & 460.6(1) & ... & 2.97(9) & $<3.2$ & 7.5(2.3) & 1.9(6) \\
 190301$^c$ & 58543 & 18:03:02.4799(4) & 459.8(4) & $-33_{-19}^{+8}$ & 2.44(8) & $<2.6$ & 3.4(1.3) & 1.4(5)
 \\\hline\hline
\end{tabular}}
\label{ta:bursts}
\end{center}
$^a$ Unconstrained parameters are listed as ``...''  Uncertainties are reported at the 68\% confidence level. Reported upper limits are at the 90\% confidence level.\\
$^b$ Bursts with multiple components have one (topocentric) arrival time and several widths {\ef and peak fluxes} reported; the arrival time refers to the first sub-burst, and width {\ef and peak flux values} for each component are presented in order of arrival.\\
$^c$ Baseband data recorded for the burst.\\
$^d$ From S/N-optimization.
\end{table}

\begin{figure}[t]
	\centering
\includegraphics[width=0.95\textwidth]{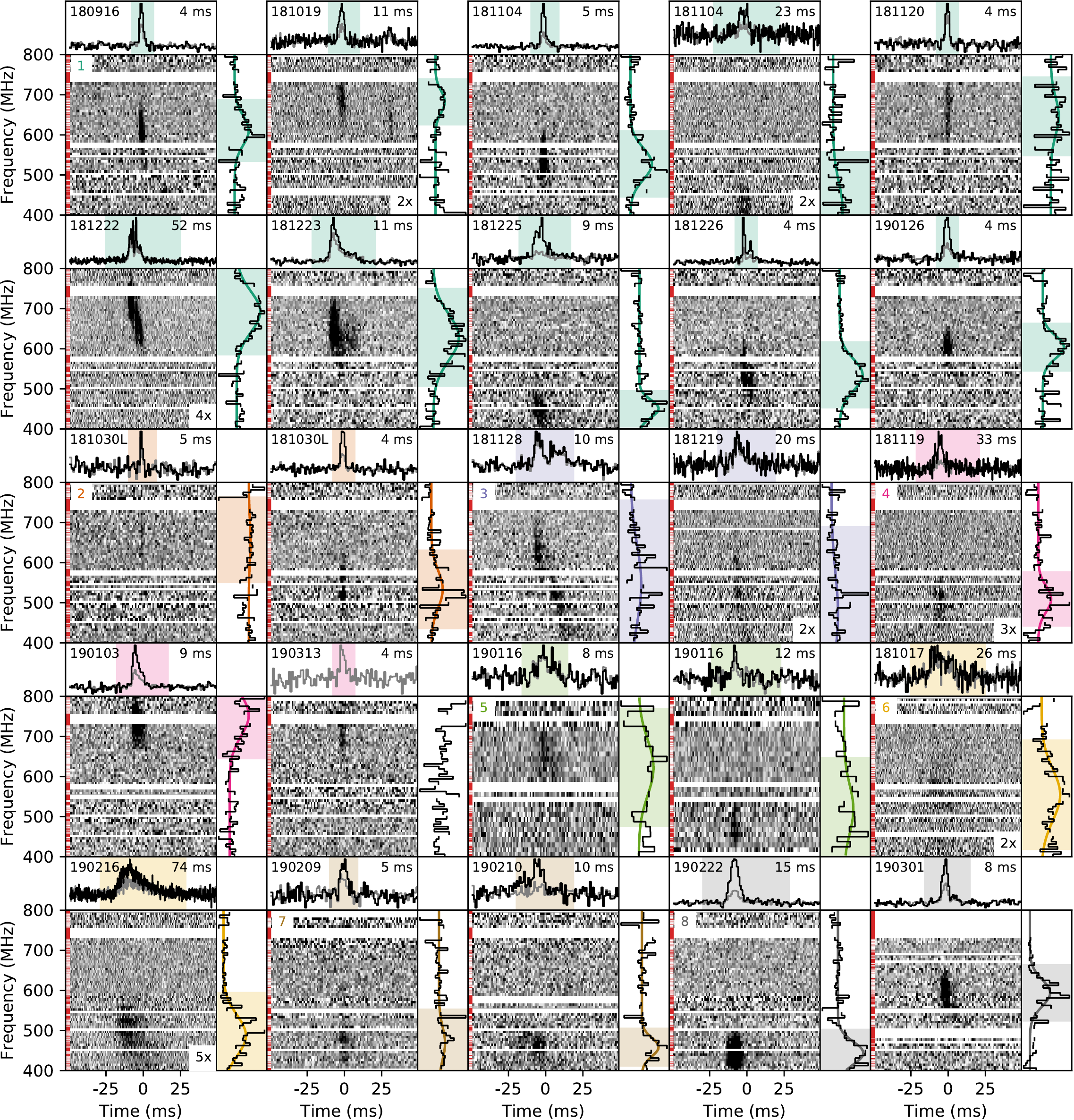}
\figcaption{Frequency {\zp versus} time (``waterfall'') plots of the bursts listed in Table~\ref{ta:bursts}, for the per-burst optimal DMs as determined in \S\ref{sec:bursts}. Every panel shows the {\zp 0.98304-ms time resolution} dedispersed intensity data with the integrated burst profile on top and the on-pulse spectrum on the right. One color is used per source.  Windows show 100 time samples ($\sim$100 ms), unless indicated otherwise by the multiplicative factor in the bottom right corner. Intensity values are saturated at the 5$^\mathrm{th}$ and 95$^\mathrm{th}$ percentiles. All bursts were detected in the source's upper transit, unless an ``L'' in the top{\zp -}left corner indicates a detection in the lower transit (only the case for Source 2). Pulse widths, defined as the width of the boxcar with the highest S/N after convolution with burst profile, are in the top right corner. The shaded region in the profile (four times the pulse width) was used for the extraction of the on-pulse spectrum. The shaded region in the on-pulse spectrum shows the full width at tenth maximum (FWTM) of a Gaussian fit, except for the third bursts of {\zp Source} 4, for which a fit did not converge. In the burst profiles, the black lines are the integration over the FWTMs and the gray lines are the integration over the full bandwidths. 64 (32) frequency subbands {\zp with a 6.25 (12.5) MHz subband bandwidth} are shown for all bursts (bursts from Source 5). Underlying missing or masked channels of the full-resolution (16,384-frequency-channel) intensity data are depicted by red lines on the left of the intensity data.}
\label{fig:waterfall}
\end{figure}


\subsection{Baseband Detection of Source 1}
\label{sec:basebandR3}

\begin{figure}[t]
\includegraphics[width=0.45\textwidth]{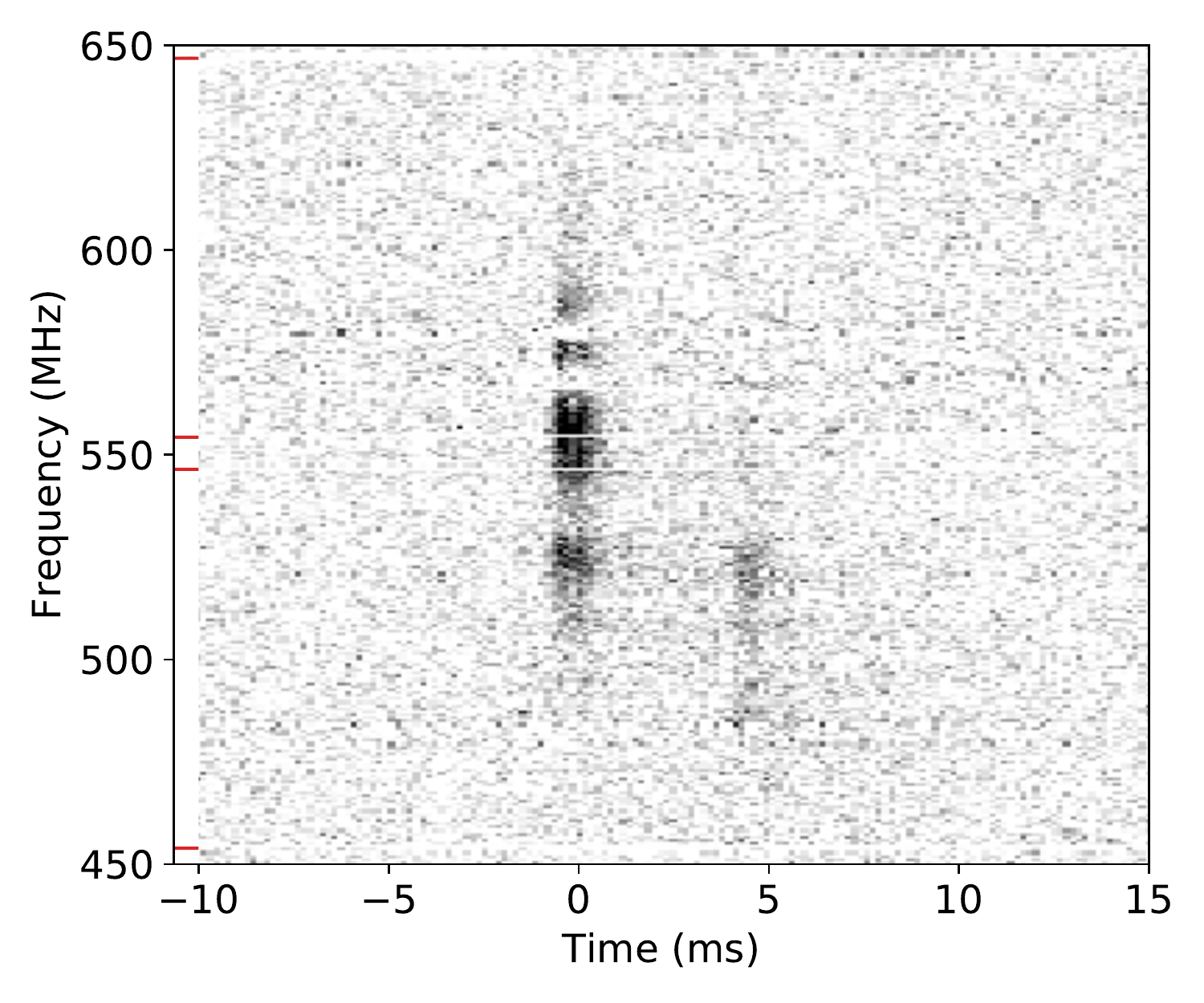}
\includegraphics[width=0.45\textwidth]{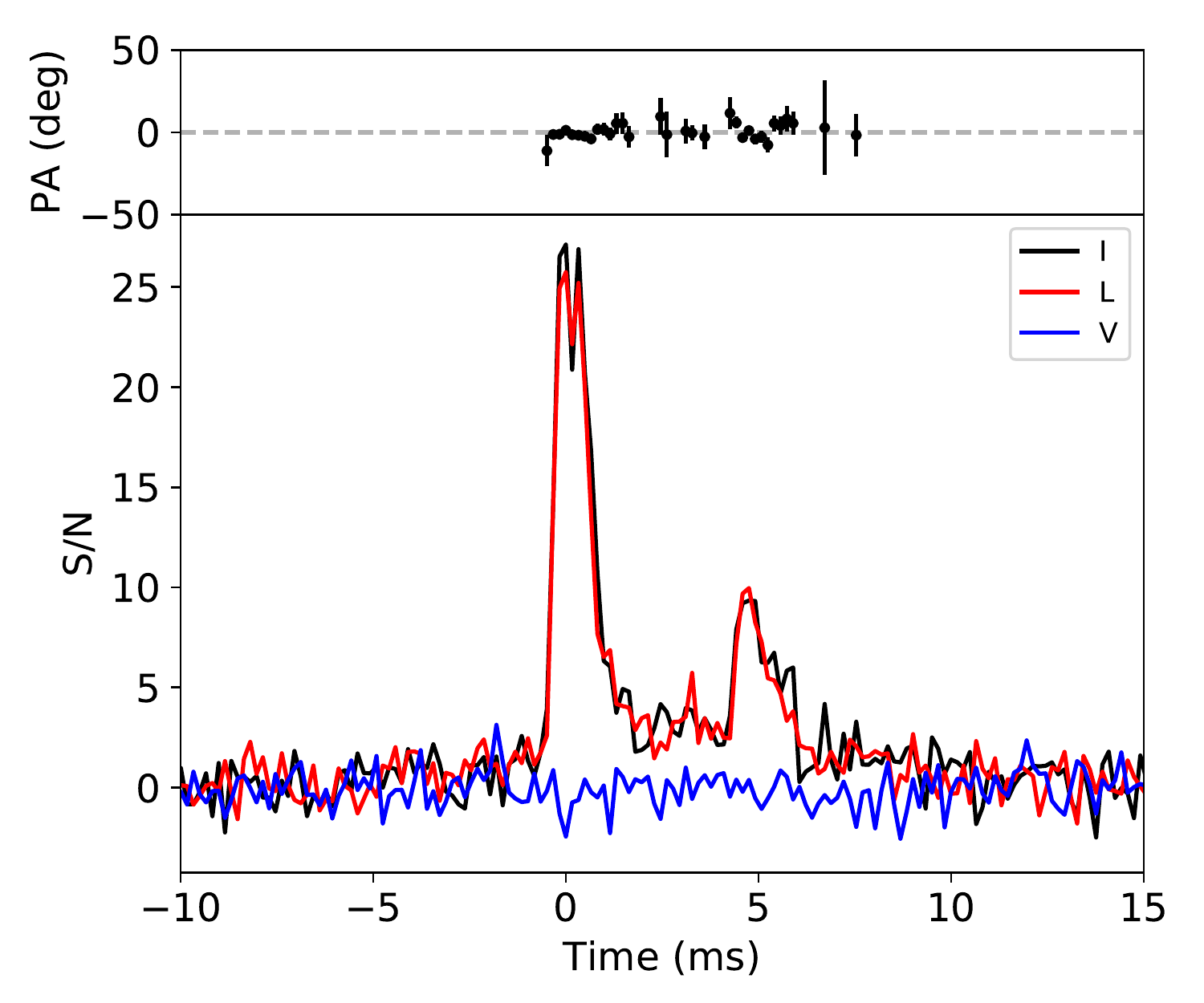}\\
\includegraphics[width=0.9\textwidth]{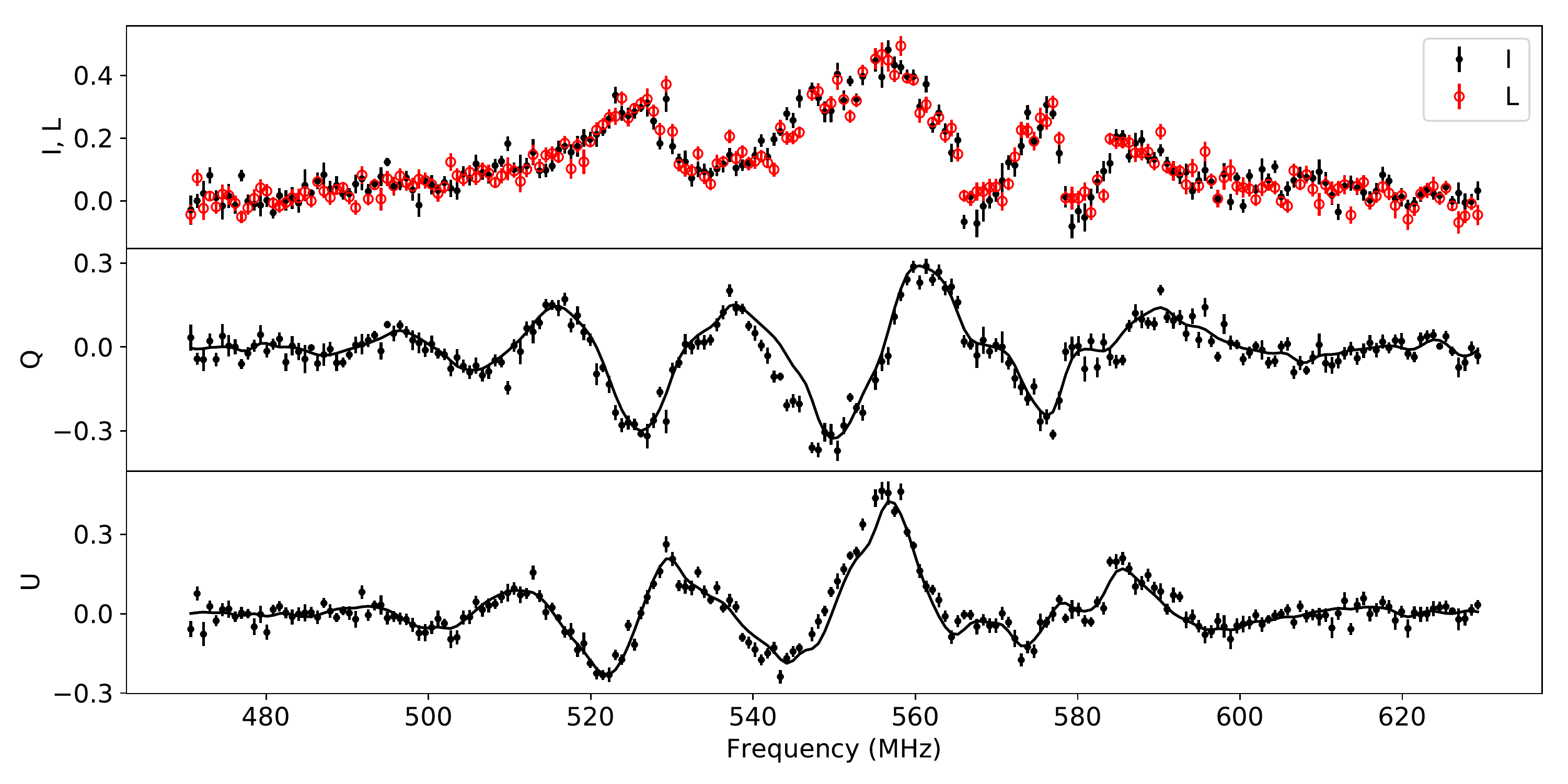}
\figcaption{Baseband data of the 181226 burst of Source 1.
    Top, left: Total intensity dedispersed ``waterfall". Red ticks indicate frequency channels masked due to RFI.
    Top, right: Pulse profile for total intensity (I, black), linear polarization (L, red) after correcting for the detected RM, circular polarization (V, blue) and the polarization position angle (upper segment).
    Profiles are averaged over frequency using S/N$^2$ weights.
    Bottom: Polarized spectra for the burst in units of S/N per frequency bin. The top panel shows data for I and L. L is obtained after correcting for the best-fit RM. The lower panels show the measured Stokes Q and U parameters. The black curve is a smoothed version of the total linear polarization data modulated in the expected way for the best-fit Faraday rotation.}
\label{fig:baseband}
\end{figure}

The new software mode to enable baseband capture was being operated occasionally during this commissioning run (see 
 \S\ref{sec:baseband}) and recorded data during an event seen from Source 1 on 181226.
A tied-array beam was formed in the direction of the best source position
calculated by the real-time pipeline. 
A total of 896 of 1024 frequency channels 
recorded dual linearly polarized voltages.  Missing frequencies are primarily due to GPU nodes that were offline at the time.  We fit the DM of the burst as in \S\ref{sec:bursts} after coherently dedispersing the data to a fiducial DM value close to the DM of the burst (to mitigate intra-channel smearing). We find $\text{DM} = 348.81 \pm 0.07$ \,pc\,cm$^{-3}$, consistent with the fit of the intensity data for the same burst, and coherently dedisperse the baseband data to this value. 
The resulting waterfall plot is presented in Figure~\ref{fig:baseband}. The higher temporal resolution of the baseband data compared with the intensity data allowed a characterization of the burst sub-structures in greater detail than possible with the intensity data. A more thorough analysis of pulse morphology/substructure will be explored in a forthcoming paper dedicated to baseband results.

Since polarization information is preserved in the baseband data, we were able to measure the polarization and Faraday rotation of the burst. \texttt{RM-tools}\footnote{\url{https://github.com/CIRADA-Tools/RM}}, a package that implements Rotation Measure Synthesis \citep{burn66,bd05}, was used for the initial rotation measure detection. The method relies on transforming polarized intensity as a function of $\lambda^2$ to Faraday depth, $\phi$, representing polarized intensity for different trial RMs. A determination of the centroid of the peak of the polarized intensity in Faraday depth space yields $\textrm{RM}= -114.6 \pm 0.6$\,rad\,m$^{-2}$. The uncertainty on this measurement was estimated in a manner analogous to radio imaging \citep{con97}, using the relation $\sigma=\textrm{FWHM}/(2\, \textrm{S/N})$, where the FWHM characterizes the width of the peak in Faraday depth space. Since we expect a single Faraday depth for FRBs, due to a compact emission region, we confirmed the RM value by using a direct model fit to the spectrum of Stokes Q and U, finding $\textrm{RM}= -115.3 \pm 1.0$\,rad\,m$^{-2}$, compatible within the uncertainties. Our expectation is corroborated by the nearly ~100\% fractional polarization of this event. An event with emission over a substantial range of Faraday depths should show variable levels of depolarization over the observing band which is not observed in 
this event. {\ef Conversely, any substantial change in the polarization position angle as a function of time should produce some degree of depolarization. This is not observed for this event, consistent with the flat position angle curve in Figure~\ref{fig:baseband}}. The difference of $0.7$\,rad\,m$^{-2}$ between the RM values obtained through the two methods is the origin of the discrepancies between the curve and data points in the two bottom panels of Figure~\ref{fig:baseband}. We ascribe this small difference to sub-dominant polarization leakages coupled with different treatments of the spectrum. Our polarization measurements are expected to be contaminated by leakages that mix the Stokes parameters, in particular the net delay between X and Y polarizations which would mix Stokes U and V (we calibrate the arrays of X and Y antennas independently). In addition, the X and Y polarizations have substantially different primary beams, and the differential response should lead to a spatially dependent leakage from Stokes I to Q. However the characteristic signature of Faraday rotation, relatively high degree of consistency of our fits, low amount of observed circular polarization, and the $\sim 100\%$ linearly polarized fraction (Figure~\ref{fig:baseband}), all indicate that these leakages are small. 

We include no correction for the ionospheric contribution to the rotation measure but expect it to be small ($\sim$1 rad m$^2$) based on preliminary ionospheric modelling. A revised RM, properly correcting for the (small) ionospheric contribution, is work in progress.

\subsection{CHIME/Pulsar Detections}
\label{sec:pulsardetection}
Out of all the repeaters reported here, the CHIME/Pulsar backend has detected only Source 1 bursts 181222, 181223, and 190126. The observations on 181222 used two independent tied-array beams, each with beam widths of $30^\prime$ at 600 MHz, which were centered on two slightly different positions following the previous CHIME/FRB detections. The burst from Source 1 was detected in both positions, at R.A., Dec. = [29.29$^\circ$, 65.79$^\circ$] with S/N of 57.1 and at R.A., Dec. = [29.16$^\circ$, 65.83$^\circ$] with S/N of 38.0. On 181223, only the first listed position (the one with higher S/N) was observed with CHIME/Pulsar, and this yielded a detection of a burst with S/N of 24.3. On 190126, a weak burst of S/N 10.3 was detected at the same position. For all the other repeaters, repeat bursts were observed by the CHIME/FRB instrument (blue dots in Figure~\ref{fig:psrfollowup}) with sufficiently high S/N that, in principle, CHIME/Pulsar could have detected them. Unfortunately, it was offline or pointed at other sky locations at these times. 
Figure~\ref{fig:psrfollowup} shows the hours of exposure for each source with CHIME/Pulsar. 
We monitored the coherently dedispersed time series, which includes instrument noise and uncorrected bandpass effects, to find the largest excursion during the observation period. After initial RFI mitigation using the {\tt rfifind} utility in \presto,  we report the brightest CHIME/Pulsar events consistent with the CHIME/FRB position and DM, and label this as the S/N upper limit from the Pulsar observations.

\begin{figure}[tb]
\includegraphics[scale=0.8]{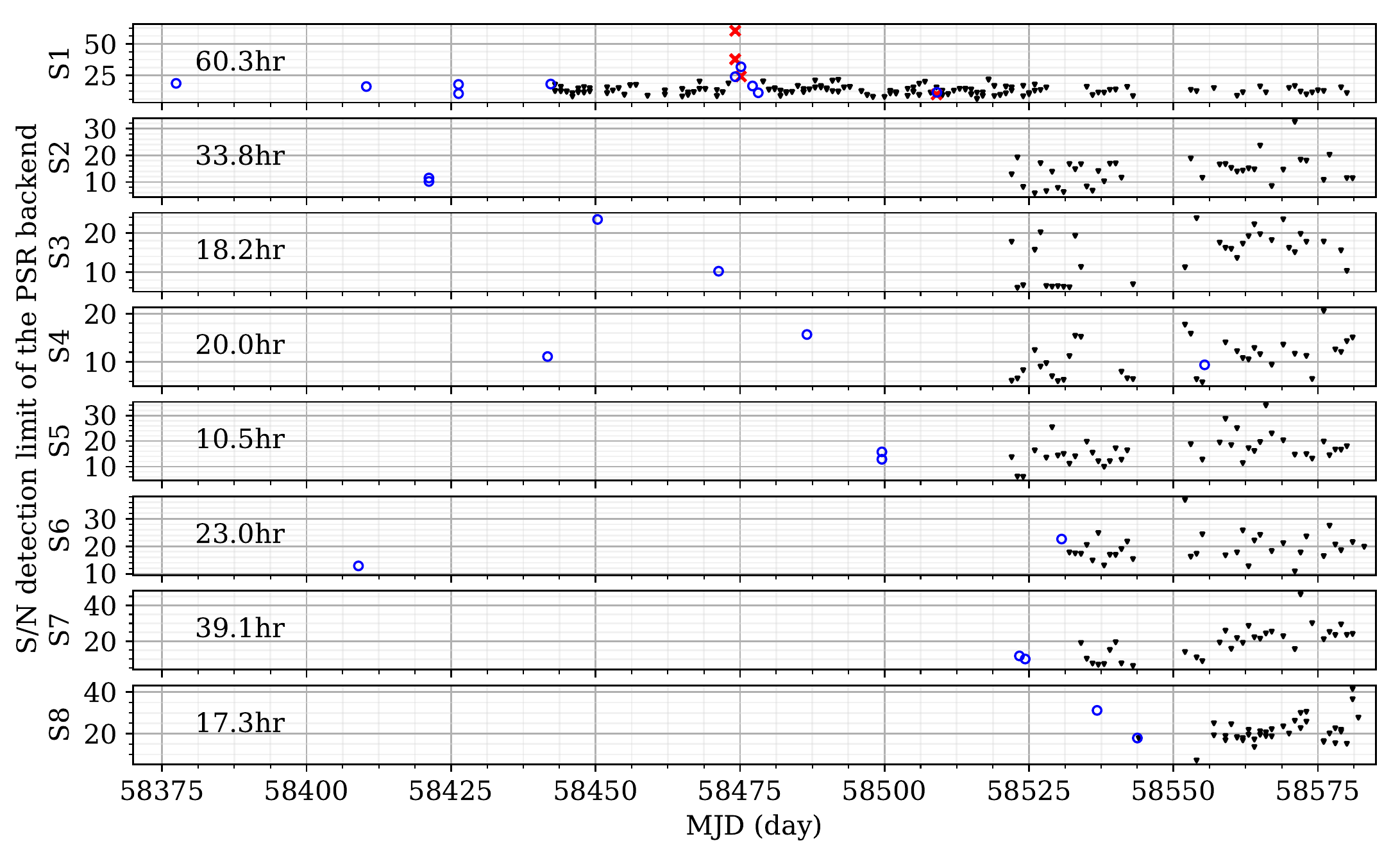}
\figcaption{Monitoring of CHIME/FRB repeaters reported here with the CHIME/Pulsar backend. Only repeating Source 1 has been detected in coherently dedispersed filterbank data taken with the CHIME/Pulsar backend (red crosses). For all the other observations in which no burst is detected, we show a detection limit in S/N as black downward triangles (see \S~\ref{sec:pulsar}.) A rigorous analysis of these data, with improved RFI-excision, is underway. The total monitoring time spent on each repeater by the CHIME/Pulsar backend is annotated on each panel in  hours. For reference, the detections from the CHIME/FRB backend are also shown as blue circles. No CHIME/Pulsar observations coincided with the FRB detections, except on three days for Source 1.  This is further discussed in Section~\ref{sec:pulsardetection}.}
\label{fig:psrfollowup}
\end{figure}

\section{Discussion}
\label{sec:discussion}

The discovery of eight new repeating FRB sources by CHIME/FRB represents important progress in the FRB field.  Although we do not yet know the origin or nature of repeaters, these new sources, along with the second identified repeater FRB 180814.J0422+73 \citep{abb+19b}, present opportunities for localizations via interferometry, and eventual host galaxy and redshift determinations.  Moreover, the study of this new population's properties, particularly in comparison with those of apparent non-repeaters, could reveal differences in emission mechanism or environment.  However, caution is required as apparent non-repeaters could later be shown to repeat \citep[see also, e.g.,][for a discussion of the observational challenges of detecting repeat bursts]{cp18, csr+19}. 

Although our current localizations are only at the precision level of {\mb $\sim$10$^\prime$}, there are already interesting inferences we can make on potential counterparts, depending on their excess DM and corresponding upper limit on redshifts. Specifically, we focus on Sources 1 and 2, which have the lowest excess DM values, then consider our repeaters as a population, accounting for the other two known repeating sources FRB 121102 \citep{sch+14,ssh+16a} and FRB 180814.J0422+73 \citep{abb+19b}.

\subsection{Source 1}
\label{sec:r3}

Inferring an extragalactic origin for Source 1 requires that the observed DM (349~pc~cm$^{-3}$; see Table~\ref{ta:bursts}) exceeds the contribution from the Milky Way and its halo along the Source 1 line-of-sight. The predicted Galactic DM contributions for this line-of-sight from the NE2001 \citep{ne2001} and YMW16 \citep{ymw17}  models are approximately 200~pc~cm$^{-3}$, and  325~pc~cm$^{-3}$, respectively. {\mb Such disagreement between the predictions of the two models is 
{\vk common}
for FRB sources {\vk like Source 1}
{\vk at} low Galactic latitude.}
If we assume a halo DM contribution of 50--80~pc~cm$^{-3}$ \citep{prochaska2019probing}, the YMW16  model  places Source 1 within the Milky  Way halo. {\mb{{\vk At} the onset, the similarity in pulse structure of Source 1 to that observed from known repeaters FRB 121102 and FRB 180814.J0422+73 suggests that Source 1 may also be 
extragalactic. 
However, 
{\vk it is not conclusively known that radio pulsars and RRATs never produce similar emission structures;}
we thus cannot establish the extragalactic nature of Source 1 on this basis alone. Here we instead use other information about Source 1 and its line-of-sight to place independent constraints on its location.}} In what follows, we consider whether Source 1 is extragalactic or Galactic based on independent estimates of the Milky Way DM and RM contributions.

\subsubsection{Different Estimates of the Milky Way DM Contribution}
\label{sec:DMestimates}

Within the Source 1 localization region, we find no catalogued Galactic ionized region that could contribute excess DM unaccounted for by the NE2001 and YMW16 models \citep{anderson2014wise, avedisova2002catalog}. 
Next, we independently estimate the DM contribution of the Milky Way along this line-of-sight by three different methods using archival data, to compare with the model values.


In the first method, we use an empirically derived $N_\mathrm{H}$-DM relation from \cite{he2013correlation} and the estimate of the Milky Way neutral hydrogen column $N_\mathrm{H}$ = 4.6 $\times$ 10$^{21}$ cm$^{-2}$ along the Source 1 line-of-sight from the HI4PI survey \citep{collaboration2016hi4pi}. This method estimates a Milky Way DM for Source 1 of 
153$^{+66}_{-46}$~pc~cm$^{-3}$.
However, we note that the \citet{he2013correlation} relation makes use primarily of nearby sources; in the outer galaxy where Source 1 was seen ($l = 129^{\circ}$), the mean \citet{he2013correlation} source distance is only 1.3 kpc.  
Whether that relation remains valid out to much large distances particularly in the outer Galaxy is not known.  Indeed the DM to $N_{\rm H}$ ratio in the halo likely rises since the ionization fraction rises, so the above DM estimate is likely a lower limit.

In a second method, we estimate the emission measure  (EM~$\equiv \int n_e^2 dl$) in two ways:  using H$\alpha$ and free-free flux densities.  We then use the EM values to estimate the range of possible maximum Galactic DMs for the Source 1 line-of-sight. For the H$\alpha$ flux density, we use data from the Virginia Tech Spectral-line Survey \citep[VTSS;][]{dennison1999virginia}.  We use the relation $\mathrm{EM}  = 2.75 \times \mathrm{T}_{4}^{0.9}\times\mathrm{I}_{\text{H}\alpha}$ \citep{pengelly1964recombination}, where the H$\alpha$ flux in this direction is  I$_\mathrm{H\alpha}=6.3\pm0.9$ rayleighs, and T$_4$ is temperature in units of 10$^{4}$ K. To correct for dust extinction, we used the IPHAS 3D 
map \citep{sale20143d} and found an extinction $\sim$8 mag using a foreground screen model for the dust distribution \citep{bannister2014galactic} that assumes maximal extinction.  Assuming $T_4 =1$, we find an extinction-corrected EM $\approx 136\pm19$~pc~cm$^{-6}$. 
Second, we used the Planck all-sky 
EM map \citep{planck2016} 
 to find EM $\approx 102 \pm 7$~pc~cm$^{-6}$ for
this line-of-sight.  
In addition, we examined data from the Canadian Galactic Plane Survey \citep{tgp+03}. These data reveal no isolated emission region of size up to about 1.5$^{\circ}$ that overlaps the position of Source 1 to a conservative upper limit of EM=5~pc~cm$^{-6}$.


We convert EM to DM using DM = $\sqrt{\text{EM}\times L \times f }$. Here, $L$ is the path length traversed by the FRB pulse in the Milky Way, and $f$ is the volume-filling factor of ionized gas. 
Using the work of
\citet{gmcm08},
we adopt a lower limit 
$f \gapp 0.04$ since $f$ is thought to increase with distance from the Plane.  We conservatively 
use 
the most distant H{\sc i}\ feature observed in 
the Effelsberg Bonn H{\sc i}\ Survey \citep{kerp2011effelsberg}
data  to estimate $L\approx$ 7 kpc. This leads to an estimate of the Milky Way's contribution to the FRB's DM $\sim 
200\,\mathrm{pc\,cm^{-3}}$ for
the H$\alpha$ EM, and $\gapp 
170\,\mathrm{pc\,cm^{-3}}$ for
the Planck EM.

The Galactic DM estimates from these alternative methods do not allow us to conclude whether Source 1 is extragalactic or Galactic, but they do tentatively suggest that
YMW16 may somewhat overestimate
the maximal DM along this line-of-sight.




\subsubsection{Is Source 1 Galactic or Extragalactic?}
\label{sec:S1galorxgal}


Here we discuss the implications of both Galactic and extragalactic scenarios for Source 1.

\paragraph{Galactic/Halo location}
If Source 1 is at the edge of the Milky Way or in its  halo
(distance $\sim$ 20--200\,kpc), it would be an unusual object. 
The surface density of radio pulsars decreases radially outward from the central region of the Milky Way and can be modelled as peaking at galactocentric radius $\sim$3~kpc and falling off at larger radii \citep[e.g.,][]{yk04,fk06},
and young pulsars and magnetars are concentrated heavily in the Galactic disk, with scale heights 0.33\,kpc \citep[e.g.,][]{lfl+06} and 0.02--0.03 kpc \citep{ok14}, respectively.
Thus the probability of finding a young, active neutron star at this sky position at the edge of the Galaxy or halo is low. However, it is not impossible given that some radio pulsars are known to have very high space velocities \citep[e.g.,][]{cvb+05}.

The distance range to Source 1, if Galactic, implies isotropic energies for its observed bursts between $10^{28-31}$ erg, intermediate between and inconsistent with those for other FRBs ($10^{38-41}$\,erg) and those of radio pulsars ($10^{23-25}$\,erg). Some Source 1 bursts could be up to 10$^2$ times brighter than those of the very bright Galactic radio pulsar PSR~B0329+54, though within the range of giant radio pulses \citep[e.g.,][]{kramer2003simultaneous}.  The downward-drifting burst substructure seen for some bursts of Source 1 (Fig.~\ref{fig:waterfall}) has thus far been associated primarily with the other two known repeating FRBs \citep{hss+18,abb+19b}. While analogous ``spectro-temporal" structure has been reported in the Crab pulsar's giant pulses \citep{he07}, it remains unclear if and how such {\vk an emission} 
is related to FRB downward-drifting structure.
 

\paragraph{Extragalactic location} 
On the other hand, if Source 1 is extragalactic, it {\mb{could}} be one of the nearest repeating FRBs.  Based on the estimated DM contribution of the Milky Way and halo discussed above, the maximum excess DM for Source 1 is $\sim$100\,$\mathrm{pc\,cm^{-3}}$, corresponding to a maximum redshift of $\sim$0.11 (distance of $\sim$500 Mpc), assuming an IGM-DM relation of DM$ \simeq 900z\;\mathrm{pc\,cm^{-3}}$ \citep{mcq14}. 

The current localization uncertainty of Source 1 does not allow a unique host identification. We searched 
the Advanced Detector Era (GLADE, v2.3) catalog of nearby galaxies \citep{dgd+18} for counterparts in the 90\% localization region (see Table~\ref{ta:repeaters} and Fig.~\ref{fig:localization}) and found five galaxies with distance $<$ 500 Mpc. The updated catalog is considered to be complete to 300\,Mpc for galaxies brighter than absolute magnitude $M_B= -20.47$\,mag \citep{arcavi2017optical}, but fainter galaxies may have been missed. 

Motivated by the presence of a persistent radio continuum source coincident with the repeating FRB 121102 \citep{clw+17}, we examined archival radio data.
We found four cataloged radio sources within Source 1's  90$\%$ localization area in the 1.4\,GHz NRAO VLA Sky Survey \citep[NVSS;][]{ccg+98}. Using the \texttt{Aegean} software package \citep{hancock2012compact,hancock2018source}, we detected five sources in the Quicklook images from the 3\,GHz VLA Sky Survey \citep[VLASS\footnote{\url{https://science.nrao.edu/science/surveys/vlass}};][]{lbc+19} and measured their integrated fluxes. Two of the NVSS objects are resolved as double-lobed extended sources in VLASS.  For the other two sources, we compared the NVSS 1.4\,GHz fluxes to the VLASS 3\,GHz fluxes and found that their spectral indices ($-0.4\pm0.1$ and $-0.8\pm0.3$) are consistent with those of radio galaxies \citep{randall2012spectral} and are different from that of the persistent source of FRB 121102 \citep{clw+17}. We found one VLASS source that is not present in NVSS but is co-located with SDSS\,J015840.07+654159.5, a galaxy with photometric redshift $z=0.39$, beyond the maximal distance we infer for Source 1 \citep{alam2015eleventh}. None of these sources are consistent with being a unresolved, flat spectral index persistent radio source similar to that co-located with FRB 121102.
Thus, if Source 1 had an unresolved persistent radio source similar to that of FRB 121102 {\mb{\citep{mph+17}}}, {\mb{its luminosity would be }} 
$\nu L_\nu<5\times10^{38}\,\mathrm{erg\,s^{-1}}$ 
$(5\sigma$) in the 2--4\,GHz frequency range, at a distance of 500 Mpc, based on the single epoch sensitivity of VLASS (120\,$\mu$Jy/beam).  This upper limit is smaller for a closer source.}  In comparison, the persistent source associated with FRB 121102 has a luminosity of $\nu L_\nu\approx7\times10^{38}\,\mathrm{erg\,s^{-1}}$ in the 2--4 GHz band \citep{clw+17}.

The short observed scattering time of Source 1 (see Table~\ref{ta:bursts}) of $\lapp 0.3$~ms at 1 GHz does not, alone, provide a strong argument for or against an extragalactic origin.
\cite{cordes2016radio} showed that the scattering times of FRBs are generally below those of Galactic radio pulsars of similar DM \citep[see also][]{abb+19a}.  This is not surprising given our disk location in the Milky Way as well as those of most Galactic radio pulsars.  Source 1, like other known FRBs, shows a scattering deficit with respect to the mean Galactic trend.  This could imply that Source 1, like other FRBs, is extragalactic.  However this is not conclusive, given that, if Source 1 is Galactic, it would be  in the halo where we have limited knowledge of scattering times, and in fact might expect reduced scattering compared to that in the Galactic disk. On the other hand, our measured scattering time of $\lapp$2~ms at 600 MHz for Source 1 implies 0.3~ms at 1 GHz, to be compared with the NE2001 predicted scattering time at 1~GHz of $>0.02$~ms.  Though not inconsistent, the larger observed value may hint at extragalactic scattering.

Strong evidence for a Galactic or extragalactic location for Source 1
could come via an interferometric localization. If Source 1 is localized to a nearby galaxy, a low chance coincidence probability could disprove a Galactic location. 
Alternatively, the proper motion of a Galactic halo object may be measurable with Very Long Baseline Interferometry (VLBI) observations over a few decades. The rotational velocity at the distance of the halo $\sim150\,\mathrm{km\,s^{-1}}$ \citep{bck14,bb16} would lead to a motion of about 1.5\,mas every 10 years. Larger proper motions would be expected for a high velocity object.

\subsubsection{Implications of the Rotation Measure of Source 1}
\label{sec:rmdisc}

Using our CHIME/FRB baseband data, we have measured an
RM for Source 1 of $-114.6 \pm 0.6$~rad~m$^{-2}$ (see \S\ref{sec:basebandR3}). 
If Source 1 is extragalactic, then the observed RM$_{\text{tot}}$ would consist of Galactic and extragalactic components, $\label{eq:RM} \mathrm{RM_{tot} = RM_{MW} + RM_{host} + RM_{source}}$, where RM$_{\text{MW}}$ includes the Milky Way and halo contributions, and RM$_{\text{host}}$ and  RM$_{\text{source}}$ are, respectively, contributions from the host galaxy and magneto-ionic plasma, either intrinsic to the FRB source or associated with its immediate environment.  Here we ignore the RM contribution 
of the intergalactic medium (IGM) which is expected to be insignificant for this nearby FRB \citep{akahori2016fast}.

The map presented by \citet{ojg+15} estimates a foreground Milky Way RM $\approx -72\pm23 \,\mathrm{rad\,m^{-2}}$ along this line-of-sight, based on Canadian Galactic Plane Survey (CGPS) and NVSS RMs.  The former are being revised, and the latter can often be unreliable at low Galactic latitudes \citep{mms+19}. \citet{obv+19}, based on the revised CGPS data, show a foreground RM $\approx -115\pm12 \,\mathrm{rad\,m^{-2}}$ in this direction, consistent with a smoothly decreasing trend in RM from longitudes 100$^{\circ}$ to 180$^{\circ}$.  This suggests the RM of Source 1 arises largely in the Galaxy, leaving only a low contribution for  RM$_{\rm host}$+RM$_{\rm source}$, certainly far lower than {\mb{~ 10$^5$ rad m$^{-2}$, the RM of FRB 121102}} \citep{msh+18}.

The low values for intrinsic RM and DM of Source 1 and its surroundings are interesting in the context of young magnetar \citep[e.g.,][]{lyu14,bel17,metzger2019fast} and supernova remnant models \citep{piro2018dispersion}. 
In the \citet{metzger2019fast} model, a magnetar wind nebula is expected to contribute significantly to the RM and DM, as well as to the persistent radio luminosity, and the values are expected to decline on a timescale of a few decades to centuries. Young supernova remnant models \citep{piro2018dispersion} predict qualitatively similar behavior. 
In the \citet{metzger2019fast} model, the lack of a bright persistent radio source (see \S\ref{sec:S1galorxgal}) would be consistent with Source 1 being old and the persistent radio source having faded.
In an alternative scenario proposed by \citet{mbm19}, a binary neutron star merger or an accretion induced collapse could also form a millisecond magnetar but with a significantly smaller ejecta mass, leading to a smaller DM and RM contribution that drops at a faster rate.   



\begin{figure}[h]
    \includegraphics[width=0.5\textwidth,height=8cm]{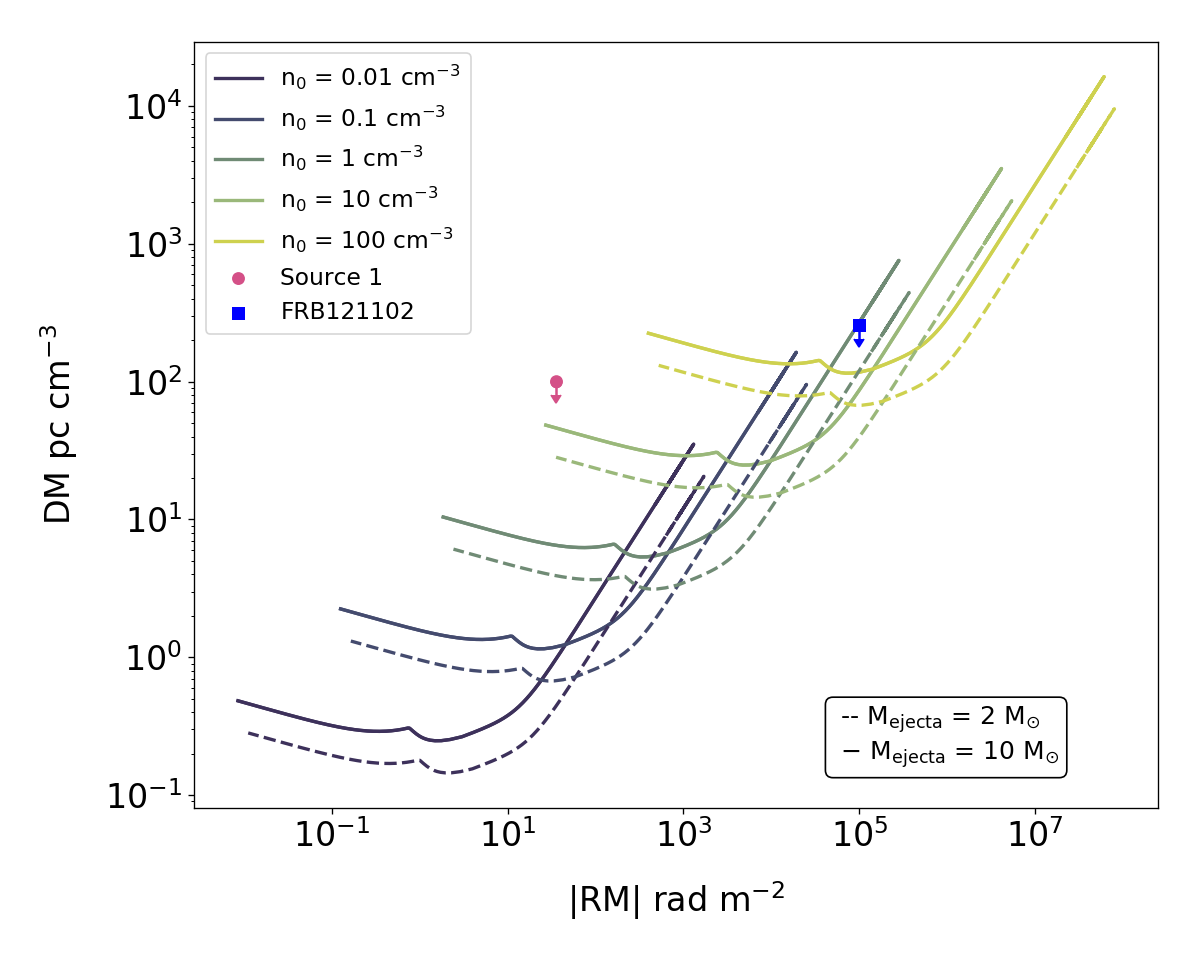}
    \includegraphics[width=0.5\textwidth,height=8cm]{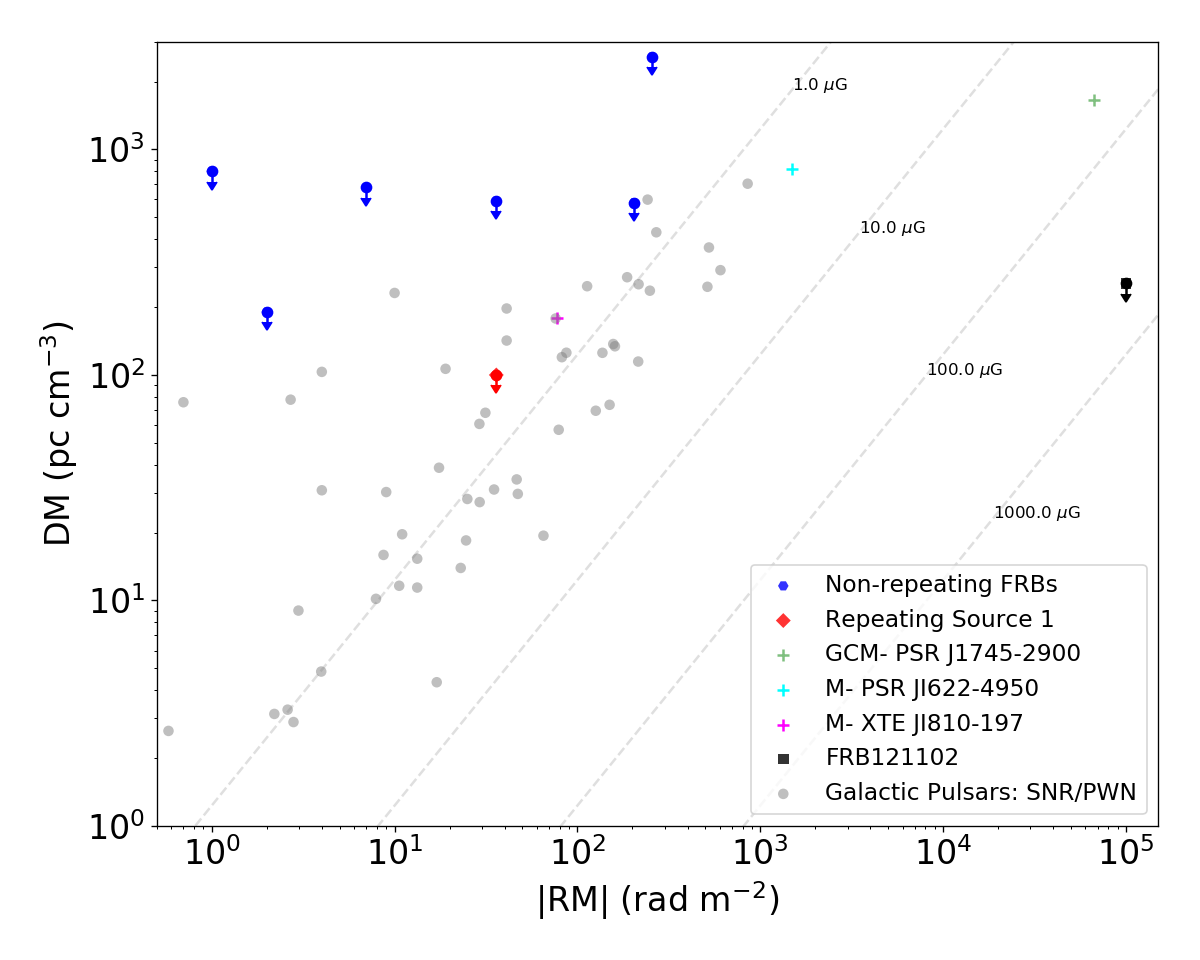}
    \caption{Left: The RM versus DM evolution for constant ISM model \citep{piro2018dispersion}. FRB\,121102 and Source 1 are shown in the plot for comparison. 
    For each ambient ISM density, the two line styles represent different ejecta masses.  Systems are predicted to evolve from right to left on a given curve. The  DM estimates shown for each FRB are upper limits since we cannot separate the host contribution from that of the medium surrounding the FRB source. For Source 1, we plot conservative upper limits for DM and RM, 100 pc cm$^{-3}$ and 36 rad m$^{-2}$ (assuming a 3$\sigma$ lower limit for the Galactic RM; see \S\ref{sec:S1galorxgal} and \S\ref{sec:rmdisc}), respectively. 
    If Source 1 is located inside supernova ejecta, either the latter are old (i.e., age $\gg 10^3 - 10^4$ yr, the Sedov-Taylor timescale) or the ambient ISM has a low number density ($\lapp$ 0.1 cm$^{-3}$). 
    Right:  Comparison between the extragalactic RMs and DMs of FRBs with Galactic radio-loud magnetars and radio pulsars that are associated with either a supernova remnant or a pulsar wind nebula. The DM values for the FRBs are upper limits as they are the total extragalactic values that include a contribution from the IGM. The Galactic magnetar and transient radio pulsar PSR\,J1745$-$2900 has an unusual line-of-sight which explains the high observed RM. Source 1, unlike FRB\,121102, lies in phase space closer to other FRBs that have not been seen to repeat thus far. 
    }
    \label{fig:SNR_RM_DM}
\end{figure}

Figure~\ref{fig:SNR_RM_DM} (left) shows the expected relation between the DM and RM contributed by a supernova remnant as predicted by \citet{piro2018dispersion}. From {\vk {F}}igure~\ref{fig:SNR_RM_DM}, we infer that if Source 1 is surrounded by supernova ejecta, it must either be old (age $\gg 10^{3}-10^{4}$ yr, the Sedov-Taylor timescale) or the ambient interstellar medium (ISM) has low density ($n_e<$ 0.1\,cm$^{-3}$).  The former scenario is disfavored because magnetar flaring activity is expected to decline rapidly with age, at least as observed for Galactic magnetars \citep{kaspi2017magnetars} while Source 1 is fairly active. 
The latter scenario disfavors models that require a dense surrounding medium like what is observed 
{\vk for} {\mb{the}} Galactic center magnetar or in the vicinity of an AGN.  


Comparing RMs and DMs of other repeating and so-far non-repeating FRBs and radio pulsars (Figure~\ref{fig:SNR_RM_DM}, right), Source 1 sits in a region occupied by other apparently non-repeating FRBs, Galactic pulsars and radio loud magnetars. 
{\vk Some}
models suggest that FRB 121102 is much younger \citep[age $<$ 100 yrs;][and references therein]{metzger2019fast,piro2018dispersion} than Galactic pulsars and magnetars. {\mb{The proximity of Source 1 to non-repeating FRBs and Galactic neutron stars in DM-RM space suggests that it may be older than FRB 121102, or that}}
{\vk Source 1 is in a very different environment.
{Some models \citep[e.g.,][]{zhang2018frb} require proximity to a supermassive black hole where a large RM is expected; this is not
as natural a location for Source 1 given its much lower RM compared to that of FRB 121102 \citep{msh+18}.}}
Moreover, some models expect the peak FRB emission frequency to decrease as the source ages \citep{lu2018radiation,metzger2019fast}, suggesting that CHIME/FRB, operating
in the 400--800~MHz band, would preferentially discover older repeaters compared to FRB 121102, which was discovered at 1.4 GHz.

\subsubsection{Implications of the Polarization Position Angle of Source 1}

{\ef Aside from the RM, the polarization position angle (PA) is another polarization product expected to be useful in elucidating possible FRB emission mechanisms. PA variations across pulse phase are often seen from pulsars, and are attributed to changes in the viewing angle geometry of magnetic field lines emanating from the polar cap of a rotating neutron star. In radio pulsars, flat PA curves can have different causes, such as originating from aligned rotators or at large emission heights \citep{phl19}. Conversely, flat PA curves may indicate an emission process not dominated by rotation \citep{michilli2018extreme}. 


Like FRB 121102, Source 1 displays a flat PA across the burst duration (see Figure~\ref{fig:baseband}). This is in contrast to FRB 110523 \citep{mls+15a}, which shows a clear PA variation across the burst duration. \citet{rsb+16} interpret this result within the context of a rotating, magnetised neutron star progenitor, finding that the observations can be described by the rotating vector model commonly used to determine pulsar emission geometries \citep{rc69}. Aside from FRBs 110523 and 121102, PA curves of most of the remaining FRBs with measured Stokes parameters have large uncertainties due to long sampling rates and large channel smearing. These difficulties are mitigated in some CHIME/FRB detections thanks to the recording of raw voltages from single antennae (see Section \S\ref{sec:baseband}). In addition to these observational limitations \citet{ckv+18} provide an additional caveat against over-interpreting flat PA curves, highlighting the flattening effect of scattering on position angle. This does not appear to be an important effect for Source 1, which has an observed scattering time that is relatively short. This is corroborated by a $\sim100\%$ linear polarization fraction across the burst duration that could not arise in the case of a strong PA swing that was subsequently flattened by scattering.} 


\subsection{Source 2}
\label{sec:r4}

Source 2 is another low-DM FRB, having DM = 103.5~pc~cm$^{-3}$, with an estimated Galactic DM contribution of 33 pc cm$^{-3}$ \citep{ymw17} to 41 pc cm$^{-3}$  \citep{ne2001} along this line-of-sight. 
In examination of archival 1420-MHz total-intensity data from DRAO Synthesis Telescope observations made in 2007 \citep[from which HI data were published by  ][]{bjf+10}, we found no evidence of any isolated  emission region of size up to $\sim$1.5$^{\circ}$ overlapping the position of Source 2 to an upper limit of EM=2~pc~cm$^{-6}$. Estimating a path length through the Local arm of 1 kpc along this line-of-sight, and adopting the method used in \S\ref{sec:DMestimates}, this translates to an upper limit on any DM contribution of local HII of only 0.3~cm~pc$^{-3}$.
Assuming the lower estimate of the Milky Way halo contribution of 50 pc cm$^{-3}$, we estimate the excess DM in this line-of-sight to be $\lapp$20 pc cm$^{-3}$. This places Source 2 at redshift $z < 0.023$  {\mb{\citep{mcq14}}}, implying a luminosity distance $<$100 Mpc. However, its current localization uncertainty region, even if assuming only the central region -- see Figure~\ref{fig:localization} -- is approximately 0.24  square degrees, too large to deduce any reliable extragalactic host association.

Nevertheless, we can make some inferences based on the low DM and making the tentative assumption that the source is most likely in its central localization region (Fig.~\ref{fig:localization}). 
We looked for nearby galaxies in different extragalactic source catalogs including GLADE 2.3 \citep{dgd+18}, which contains all of the brightest galaxies within 100 Mpc luminosity distance \citep{arcavi2017optical}. 
Notably, NGC 3403, a spiral galaxy with baryonic mass
$10^{10}$~M$_{\odot}$
with a star formation rate of $0.23\pm0.09\,\mathrm{M_\odot yr^{-1}}$ \citep{erroz2016non} at 22.2 Mpc ($z = 0.0042$) is within the localization region of Source 2. 
If NGC 3403 is its host galaxy, then DM$_{\rm host}$ +DM$_{\rm source} \leq$ 10 pc cm$^{-3}$, suggesting that the FRB progenitor is situated either at the outskirts of the NGC~3403 disk or in its halo.


We also examined different radio surveys and archival data for the presence of a FRB\,121102-like persistent radio source within
the 
localization box for Source 2.
We analyzed the Quicklook images from VLASS \citep{lbc+19} with \texttt{Aegean} \citep{hancock2012compact,hancock2018source}. After removing imaging artifacts, we found 16 radio sources in the region. 
We searched the NVSS \citep{ccg+98}, WENSS \citep{rengelink1997westerbork} and TGSS \citep{intema2017gmrt} catalogs for counterparts of these sources and possible long-duration transients such as the one decribed by \citet{lgm+18}. We found three catalogued sources that are not present in VLASS, one of them being NGC 3403. However, comparing the flux densities and sizes of these three sources in the different catalogs indicates that the three are likely resolved out in VLASS and hence not point sources. 
To determine whether any of the 16 VLASS sources could be an FRB~121102-like radio counterpart, 
we examined Pan-STARRS data at all 16 locations.
In all cases where there was a coincident
optical galaxy, 
based on the WISE color classifications \citep{wright2010wide} for AGN \citep{mateos2012using} and active starburst galaxies \citep{caccianiga2015wise}, the source is likely to be an AGN.  For the seven VLASS positions at which there was no optical galaxy in Pan-STARRS, 
we presume the radio source is likely a high-redshift radio galaxy.  At the maximal Source 2 redshift $z=0.023$, Pan-STARRS is complete for dwarf galaxies having luminosity ten times fainter than that of the FRB 121102 host \citep{chambers2016pan}.  Since we detected either no host, or only AGN galaxy hosts coincident with the 16 VLASS sources, we conclude that
no VLASS source in the main error region for Source 2 is a FRB 121102-like compact persistent radio source, and set an upper limit on such
emission of
(5$\sigma$) 0.6 mJy at 1.4 GHz.  However, the above reasoning would not hold if the true position of Source 2 were in a CHIME sidelobe (see Fig.~\ref{fig:localization}).

\subsection{Implications for the Millisecond Magnetar Model}
The millisecond magnetar model \citep{metzger2019fast} was developed to explain the properties of the first repeater FRB 121102, {\mb{its location
in a star-forming region of a low-metallicity dwarf galaxy,}}
the presence of a bright persistent radio source and the very large RM. The lack of persistent radio emission for Sources 1 and 2 as well as for CHIME/FRB discovered repeater FRB 180814.J0422+73 \citep{abb+19b}, with luminosities $\sim10$--30 times fainter than that for FRB 121102, and the low RM of Source 1 (see \S\ref{sec:baseband}) are notable. In the context of the millisecond magnetar model, the differences might indicate that these CHIME/FRB repeaters are significantly older than FRB 121102, as their persistent radio emission might have faded and the RM-producing nebula has dissipated.  On the other hand, the millisecond magnetar model may not be applicable to them.  These data may be consistent with the scenario proposed by \citet{mbm19} of a millisecond magnetar formed from a binary neutron star merger or an accretion induced collapse of a white dwarf.

\subsection{Repeater Burst Morphologies}
\label{sec:morphologies}


Fine spectro-temporal structure has been observed in a variety of (so-far-non-repeating) FRBs \citep{mls+15a,rsb+16,ffb18,msb+19} as well as in repeating FRBs {\zp \citep{ssh+16a,msh+18,abb+19b}}, and sometimes only reveals itself in baseband data. The observed structure is shaped by an unknown emission mechanism and transformed by propagation through an ionized and inhomogeneous medium -- causing scattering, scintillation and potentially plasma lensing, which has been proposed as a possible way to boost FRB signals to make them detectable at extragalactic distances \citep{cwh+17} -- and by the instrumental response of the telescope. In order to access the information that the burst structure carries, these effects need to be disentangled. Here, we investigate whether burst morphology can be used as a way of discriminating between bursts from repeating and non-repeating FRBs without necessarily having to wait for a repeat burst.

One challenge is that, without an exact model for FRB burst emission and propagation, there is an ambiguity in choosing a metric for DM optimization, especially for bursts comprised of multiple sub-bursts \citep[see also][]{hss+18}. Assuming that each (sub-)burst is emitted at the same time at all emission frequencies and that all of the radio burst frequency-dependent arrival-time delay is caused by dispersion in the ionized interstellar and intergalactic medium and not also by the emission process, we chose to optimize DMs for structure, or sharpness, of the bursts. This method leads to less scatter in the measured DMs than for conventional S/N optimization {\zp \citep[as was also demonstrated for FRB 121102;][]{hss+18}}: comparing the 23 bursts for which we were able to measure both a S/N-optimized and a structure-optimized DM, and calculating the root-mean-square deviation of the DMs with respect to the per-source average DM for that method, we find that the scatter is 0.03~pc cm$^{-3}$ for the structure-optimization versus 0.59~pc~cm$^{-3}$ for S/N-optimization. We also note that optimizing for S/N systematically estimates higher DMs than in the structure-optimizing method, especially for bursts with clearly defined structure (e.g., burst 181222 of Source 1): the S/N-optimized DM value is higher for all bursts except the 181120 burst of Source 1 (where the difference is only 0.2 pc cm$^{-3}$). Taking the residual fractional difference between each pair of measurements\footnote{We consider the fractional difference instead of the absolute difference to mitigate DM-dependent effects such as intra-channel dispersion smearing.} we find that on average the S/N-optimization leads to 0.2\% higher DM estimates.

The 2-D auto-correlation analysis, as used in \cite{hss+18} and \cite{jcf+19}, provides additional insight into the drifting structure of pulses from repeating FRBs. In several cases, the downward-drifting ``sad trombone'' observed in previous work appears to be present, and is well fit by the auto-correlation analysis. However, actual measurements are possible for only $\sim$45\% of the bursts in Figure~\ref{fig:waterfall} and for none of the events from Sources 2 and 5. Of the bursts with no measurable drift rates, more than half do not visually show downward-drifting structure; they appear to be sharp pulses. {\zp Note here that many bursts from FRB 121102 are also single-peaked and show no downward-drifting structure.} The remaining bursts have S/N approximately ten or less, and the signals are not strong {\zp enough} to yield clear results\footnote{{For some bursts, a fit to the data might initially yield a drift rate measurement, however, the Monte Carlo resampling of the DM uncertainty and noise distributions in the full auto-correlation analysis will show if that measurement is significant or not. Imposing an (arbitrary) cutoff in the S/N of bursts to analyze is thus not required.} } \citep[see also][for a {\zp comparison of low S/N bursts from FRB 121102 with a high S/N clearly downward-drifting burst with noise added until the S/N matches that of the low S/N bursts}]{gms+19}. 

These results provide an important caution in our analysis of repeating FRBs. While downward-drifting structure can be a feature of repeating FRBs and may be more common among repeaters than among non-repeating FRBs, many repeat bursts occur without any measurable downward-drifting structure. Therefore, we cannot definitively say that an observed FRB will not repeat because it lacks downward-drifting structure. 
Also, downward-drifting structure may be unresolved in the coarse time resolution of CHIME/FRB intensity data (cf. the
181226 burst from Source 1 in Figures~\ref{fig:waterfall} and \ref{fig:baseband}). Future analysis of baseband data of repeater and other bursts will be paramount to our understanding, and it may yet be the case that the presence of downward-drifting structure is predictive of repetition.

Sub-bursts marching down in frequency have now been detected in bursts from nine repeating FRBs \citep[FRB 121102, FRB 180814.J0422+73 and Sources 1, 3, 4, 5, 6, 7 and 8 in this paper;][]{gsp+18,hss+18,abb+19b} and not in any (so-far) non-repeating FRBs. All linear drift rates measured in the CHIME band are on the order few to tens of MHz/ms, though CHIME/FRB is sensitive to linear drift rates up to $\sim-$400 MHz/ms (but note that \emph{any} drift rate will be undetectable if it occurs within one $\sim$1-ms time sample). The minimum measurable drift rate depends on the frequency channel bandwidth of the instrument, as well as on the number of burst components and their S/N, but it is also related to the more conceptual question of when two bursts still belong to the same envelope. For example, for the 181019 burst of Source 1, with sub-bursts separated by $\sim$50 MHz and $\sim$60 ms, similar to the extent of the 181222 burst envelope from the same source, the drift-rate measurement is not sensitive to the second `blob' of emission and one can argue whether it is a separate burst or is part of the same burst train, given the sub-bursts' precise match in DM. 
Along the same lines one could argue that the two sub-bursts in the 181128 event of Source 3 are not part of the same envelope because they do not line up with the same DM value, although we do measure a linear drift rate for that burst and the burst separation is small ($\sim$few ms).

The drifting behavior might be emission originating from different heights in a plasma with a gradient of properties \citep[e.g.,][]{mms19,wzc19}.
Other known radio bursts produced by coherent emission, albeit {\zp (likely)} with orders-of-magnitude lower magnetic-field strengths -- pulsars, Jovian decametric bursts and Solar bursts -- show similar phenomenology in their dynamic spectra \citep{he07,e69,m17}, including drifting bursts with a trend in drift-rate evolution \citep[see e.g.,][]{e82,rzh14,bbg98}. In the framework of plasma lensing \citep{cwh+17}, bursts are expected to drift both up and down in frequency, but so far only unidirectional, downward, drifts have been observed.


The emission bandwidths of our repeater bursts (Figure~\ref{fig:waterfall}), as well as in the CHIME/FRB detections of FRB 121102 \citep{jcf+19} and FRB 180814.J0422+73 \citep{abb+19b}, tend to be only 100--150 MHz, and appear to be well described by a Gaussian. 
{\zp In} contrast, at least six of the so-far non-repeaters detected by CHIME/FRB \citep{abb+19a} extend across CHIME's full 400-MHz band and their spectra would likely be better described by a power law. {\zp FRB 121102 bursts at 1.4 GHz also tend to be narrow-band, with emission bandwidths of about 100--300 MHz \citep[FWHM;][]{gms+19}.}

However, there are two important caveats that deter us from concluding that repeater bursts have smaller emission bandwidths. First, the two samples were detected at different stages of commissioning and were thus likely subject to different selection biases. For example, the CHIME/FRB real-time detection pipeline searched over a flat burst spectral index at the start of commissioning but more recently has started to search over spectral indices $-$3 and +3, while still being nearly optimal for flat-spectrum bursts. This may have led to increased sensitivity to narrower-bandwidth bursts. Second, CHIME/FRB's beam bandpasses vary with angle away from the beam centers and especially in the sidelobes, where they will often be sensitive to only a part of the total bandwidth. This beam effect can only be properly corrected for when the exact sky position of a source is known and can otherwise lead to a degeneracy between intrinsic emission bandwidth and sky position. An analysis of a larger sample of CHIME/FRB-detected FRBs, including forward-modeling of the instrumental response and a comparison of emission bandwidths and fluences, is necessary to establish whether sources of repeating FRBs have a distinct distribution of emission bandwidths.




\subsection{Repeating FRB DMs, Burst Widths\ef{, and Peak Fluxes}}
\label{sec:widths}

If repeaters represent a population of FRBs distinct from a non-repeating population, we might expect the two to have different DM distributions, if, e.g., their host or close source environments differed, or if one population were intrinsically more luminous.  For this reason, we compared the DMs of the CHIME/FRB repeaters (the 8 reported here and FRB 180814.J0422+73) with those of the 12 apparent non-repeaters from \citet{abb+19a}.  We compare CHIME/FRB repeater DMs with other CHIME/FRB events -- and not with the DM distributions from other FRB surveys -- to ensure that selection biases are not an issue.  We found no statistically significant difference using either a 2-sided Kolmogorov-Smirnov test \citep{m51} or a 2-sample Anderson-Darling test \citep{ss87}. 
This suggests that the environments and luminosities of repeater bursts may not differ greatly from those of apparent non-repeaters, although larger samples are required to confirm this, and one or more of the 12 apparent non-repeaters may yet repeat.

Another property of FRBs which 
may observationally distinguish repeating and so-far non-repeating sources is their intrinsic temporal pulse width.
This was first noticed by \citet{ssh+16b} in a study of twelve FRB 121102 bursts for which the intrinsic widths were significantly longer than those of thirteen single-component, non-repeating FRBs detected with the Parkes telescope. Although there is a much larger sample of both non-repeating and repeating FRBs that can now be included in such a comparison, we restrict our sample to detections in the 400--800 MHz-frequency range as the variation of intrinsic width with observing frequency is not well understood. In addition to widths for the eight sources presented in this work (see Table \ref{ta:bursts}), we include widths for twelve non-repeating sources reported in \citet{abb+19a}, 
and previously known repeaters, FRB 121102 and FRB 180814.J0422+73 \citep{jcf+19,abb+19b}. For the sources which do not have a significant width measurement, we assume the corresponding 95\% confidence upper limit to be the measured value. 

\begin{figure}[h]
	\begin{center}
		\includegraphics[width=0.85\textwidth]{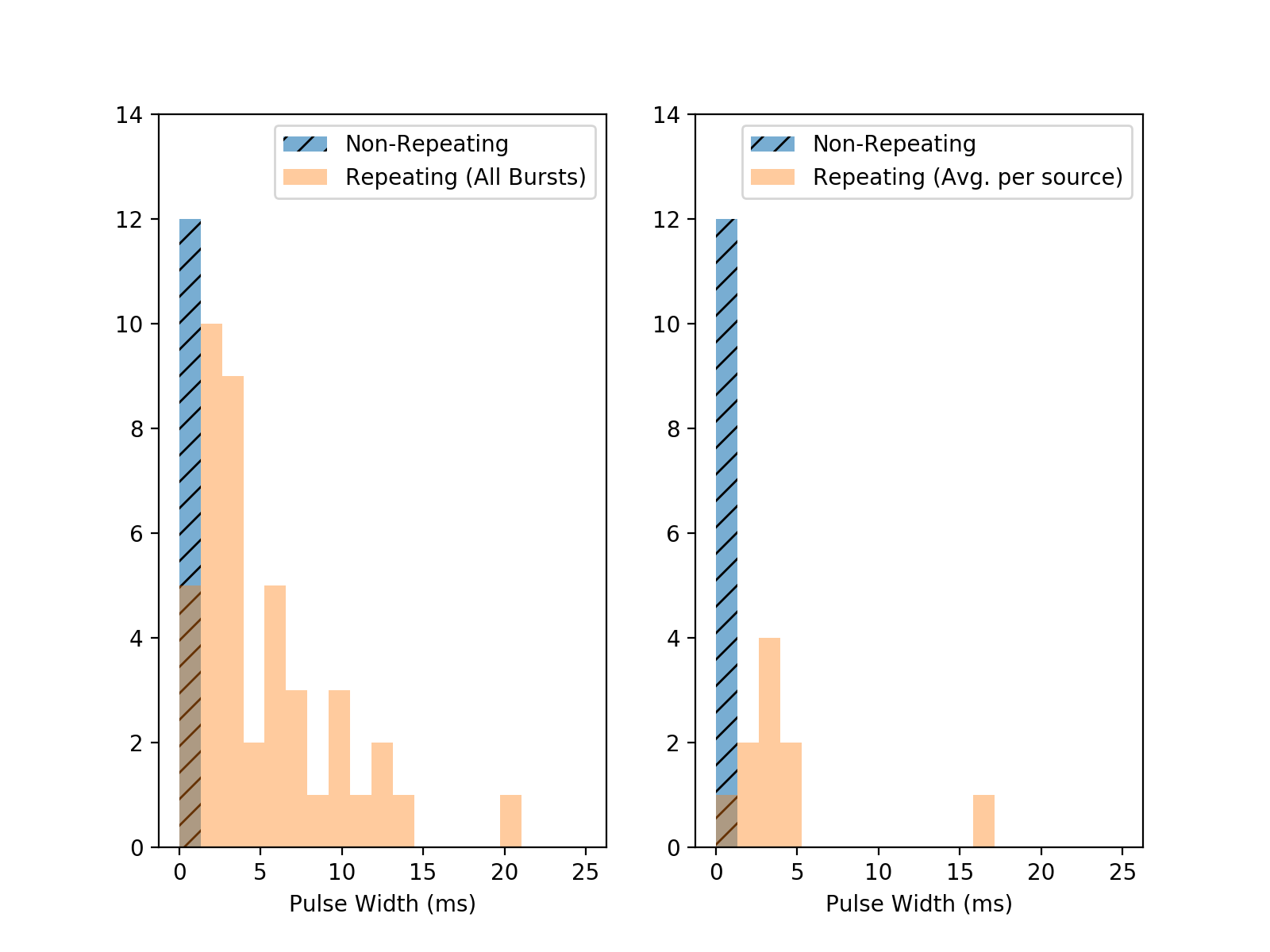}
	\end{center}
\figcaption{Distribution of intrinsic temporal widths for repeating and non-repeating FRB sources observed in the frequency range of 400--800 MHz. For repeating FRBs, the left panel shows the distribution of widths of the Gaussian spectral components for all bursts from each source while the right panel shows only the weighted average of the widths for each source.}
\label{fig:widths}
\end{figure}

{\PC Due to the reduced S/N threshold for saving intensity data for repeat bursts, the real-time detection pipeline could potentially be more sensitive to large pulse widths for bursts from repeating FRBs as compared to bursts of similar fluence from non-repeating sources. To ensure this bias is not an issue, we exclude repeat bursts with a S/N $<$ 10 (the detection threshold for new sources) from this analysis. The distribution of the widths used in the comparison is plotted in Figure \ref{fig:widths}.}

We compare the temporal widths of different Gaussian components of bursts from the repeating sources with those of the so-far non-repeating ones using a (non-parametric) k-sample Anderson-Darling test and find with $\sim$4$\sigma$ significance that these two samples are not drawn from the same distribution. Averaging the measurements for bursts from each repeating FRB source using inverse-variance weighting also yields a distribution which is inconsistent with that for the non-repeating FRBs at the $\sim$3.5$\sigma$ level. A two-sample Kolmogorov-Smirnov test confirms these results with significance $\sim$5$\sigma$ and $\sim$4$\sigma$, respectively, for the two cases.

We have verified that the apparently narrower emission bandwidths of the repeater bursts do not influence the width measurements (through, for example, hindering measurement of scattering) by refitting widths for the so-far non-repeating CHIME/FRB events \citep{abb+19a}, but using only $\sim$100-MHz portions of data.  This analysis yields no difference in results for the fitted widths, supporting the conjecture that the difference is real.  Moreover, as described above, we find no evidence for the repeater DM distribution being different from that of the apparent non-repeating CHIME/FRB published events, which rules out, e.g., a bias due to enhanced {\PC intra-channel} DM smearing.

\begin{figure}[h]
	\begin{center}
		\includegraphics[width=0.65\textwidth]{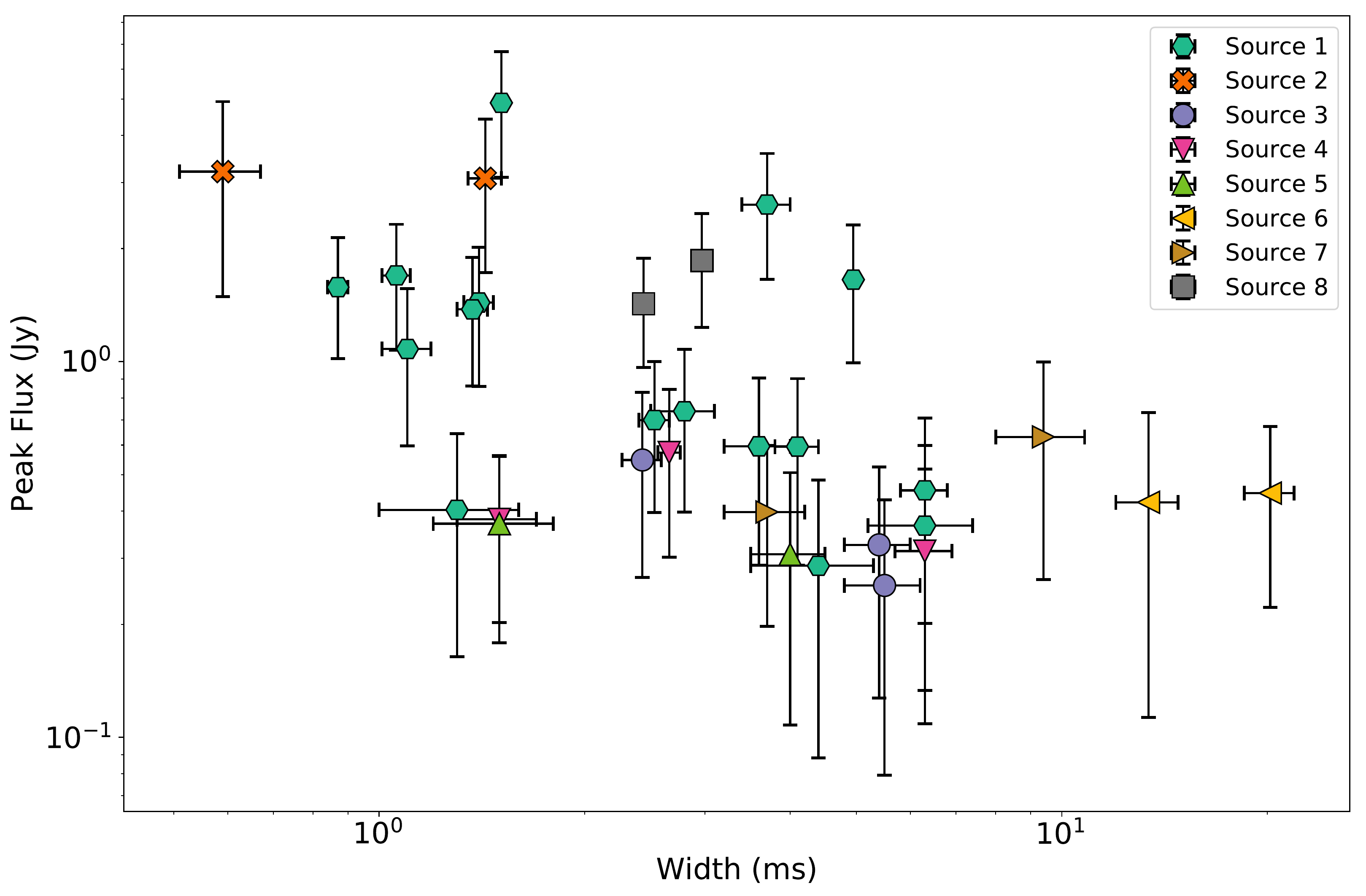}
	\end{center}
\figcaption{{\ef Peak flux versus intrinsic temporal width for each of the bursts, displayed in log space. Bursts with multiple components are represented by multiple data points, one for each sub-burst.}}
\label{fig:peak_flux_vs_width}
\end{figure}

{\ef We also compare the peak fluxes and intrinsic temporal widths of the sub-bursts in our sample, as shown in Figure~\ref{fig:peak_flux_vs_width}. To check for a correlation, we run a Monte Carlo jackknife simulation which resamples the data according to the uncertainties to obtain a distribution of correlation coefficients and corresponding p-values. In linear space, we find a mean Spearman coefficient of $\sim-0.5$ with $93\%$ of p-values indicating a greater than $2\sigma$ negative monotonic correlation. In log space, we find a mean Pearson coefficient of $\sim-0.5$ with $95\%$ of p-values indicating a greater than $2\sigma$ negative log-log correlation. Although a potential correlation would be physically interesting, we have not accounted for selection effects in our pipeline and RFI removal method. Moreover, it is unclear what biases are introduced from effects inherent to the emission mechanism that produces complex repeater morphologies. The presence of sub-burst measurements in our sample -- which likely possess correlated noise properties for sub-bursts grouped in the same detection event -- further complicates the use of the statistical tests that we described above, which assume independent and randomly sampled data. Given these circumstances, we therefore draw no conclusion despite finding marginal evidence for correlation between the peak flux and width. We will re-examine this trend once a larger FRB sample is obtained, and once our detection biases are quantified.}

\subsection{Scattering Times}


The scattering timescales reported for sources in \citet{abb+19a} suggested that CHIME FRBs are located in environments having stronger scattering properties than the quiescent diffuse ISM. This conclusion was derived from simulations modeling the dispersion and scattering properties of the FRB host galaxies, the IGM, 
 and the Milky Way. Here we perform a similar analysis to verify if the same conclusion holds for the repeating FRBs reported here.
For each FRB in our sample, we use only the strongest constraint on scattering time 
(see Table \ref{ta:bursts}). This 
assumes that scattering structures along the line-of-sight to these FRBs do not change significantly in the interval between detections. The resulting sample has one statistically significant measurement, that for Source 6, and 95\% confidence upper limits for the seven other sources,
as well as for the previously known repeaters, FRB 121102 \citep{jcf+19} and FRB 180814.J0422+73 \citep{abb+19b}.

To determine whether a population model of FRBs in a Milky-Way-like galaxy can reproduce the sample properties, we perform 50,000 simulation runs. For each of these runs, we generate DMs and scattering times at 600 MHz for the ten repeating FRBs. A run is classified as successful if at least one of the ten simulated sources has a scattering time at 600 MHz $>$  4.7 ms, which is the strongest upper limit on scattering time for the repeating FRBs reported here.  Even though there is only one such source in our observed sample, we allow for more than one simulated source to meet this criterion to account for the bias our search pipeline has against detection of highly scattered pulses \citep{abb+19a}. Based on the fraction of successful runs, we find that we cannot rule out a population of isolated FRBs located in random locations in Milky-Way-like galaxies. 


As an additional check, we compare the distribution of scattering timescales of repeating FRB sources with that of the so-far non-repeating sources presented in \citet{abb+19a}. The significant fraction of repeating FRBs having only reported upper limits makes a direct comparison of scattering times using the Anderson-Darling test (as done for the pulse widths in Section \ref{sec:discussion}) 
difficult.
Instead, we use the reported measurements and their corresponding uncertainties or the 95\% upper limits (whichever is applicable) as parameters of a normal probability density distribution for each repeating source. We then add the probability density distributions and integrate over the resulting distribution to get a cumulative density function for the scattering times of the repeating FRB population. We perform a Kolmogorov-Smirnov test between the cumulative density function and scattering measurements of the non-repeating sources. We find that the hypothesis of the scattering measurements of the non-repeating and repeating sources being drawn from the same distribution cannot be ruled out (not even at the 2$\sigma$ level) which is consistent with the results of the scattering simulations described above.

One explanation for our success in measuring scattering
times for the apparent non-repeaters \citep{abb+19a} while measuring many
upper limits for the repeaters reported on here
 is that the latter appear to have broader widths (see \S\ref{sec:widths}), which
makes the detection of short scattering times, like
those reported in \citet{abb+19a},
more difficult.

\subsection{Repetition Rates}
\label{sec:rates}

Recently, ASKAP detected 20 FRBs, none of which were observed to repeat during their reported total time periods 7.7-45.7 days at fluence sensitivity S = 26 Jy ms.\footnote{CHIME/FRB recently detected a burst that is consistent in position and DM with one of the original ASKAP sources \citep{pat19}. The repeating nature of this ASKAP burst was first discovered by \citet{kso+19}, but the parameters needed for repetition-rate calculations are not yet available. We therefore exclude this repeating source from the calculations in Section \ref{sec:rates}.} To qualitatively assess whether CHIME's repeater detections are consistent with ASKAP's non-detections, we proceed as follows.

We calculate the effective Poisson repetition rates of the repeaters that we have detected based on the effective exposure time, sensitivity and number of bursts as calculated above, but assuming nothing about the population from which each object is drawn.  We infer 68\% confidence intervals on each rate using the \citet{kbn91}, formalism.  We scale the observed rates to a detection sensitivity of S$_0$ = 1 Jy ms using a factor of (S/S$_0)^{1.5}$, where S is the sensitivity limit calculated above for each repeating source.  If FRB rates are not strongly frequency dependent, we can scale the 68\% upper limits on the repetition rates for the ASKAP FRBs (given that only one burst was observed in the given period) by a factor of (S/S$_0)^{1.5}$.  Figure 8 shows the observed and scaled Poisson repetition rates (black circles and red triangles, respectively) for the CHIME/FRB repeaters. The gray lines indicate the 68\% upper limits on the repetition rates for the individual ASKAP FRBs. We note that the  rates from CHIME/FRB are slightly inconsistent with the ASKAP upper limits on repetition.   However, this inconsistency may be because CHIME/FRB
per-object rate estimates are biased high, both compared to the true underlying rate for each source and relative to the overall underlying population of repeaters. This is because we have selected objects that have been observed to repeat at least once, from a population of similar sources that have not yet been observed to repeat. Since most of our objects have only been observed to have repeated once, the detected repeaters are very likely to be the tail of a distribution of observations of infrequent repeaters, which can produce a significant upward bias in the per-object rates relative to the true repetition rates for these sources and relative to the wider repeater population. ASKAP upper limits would not be subject to such a bias, since they are derived from an absence of repeat bursts.

The arrival times of bursts from FRB~121102 are known to be inconsistent with being derived from a Poissonian distribution at a fixed rate \citep{ssh+16a,ssh+16b,oyp18}. A few of the sources here appear to display hints of clustering:  four bursts from Source 1 were detected within five days  at  the  end  of  2018  and  two  bursts  from  Sources  2  and  5  were  detected  in  a  single  transit with no other bursts detected despite daily observations.  It is tempting to view these as evidence for non-Poissonian repetition. An analysis could be performed to determine the probability that such clusters could arise from a Poisson process. However, we caution that these repeating sources are derived from a much larger set of detected FRBs, some, or potentially all, of which may be capable of repeating. Such analyses require knowledge of the broader population of CHIME/FRB-detected events, including apparent non-repeaters, and are beyond the scope of this paper.


\begin{figure}[h]
	\begin{center}
		\includegraphics{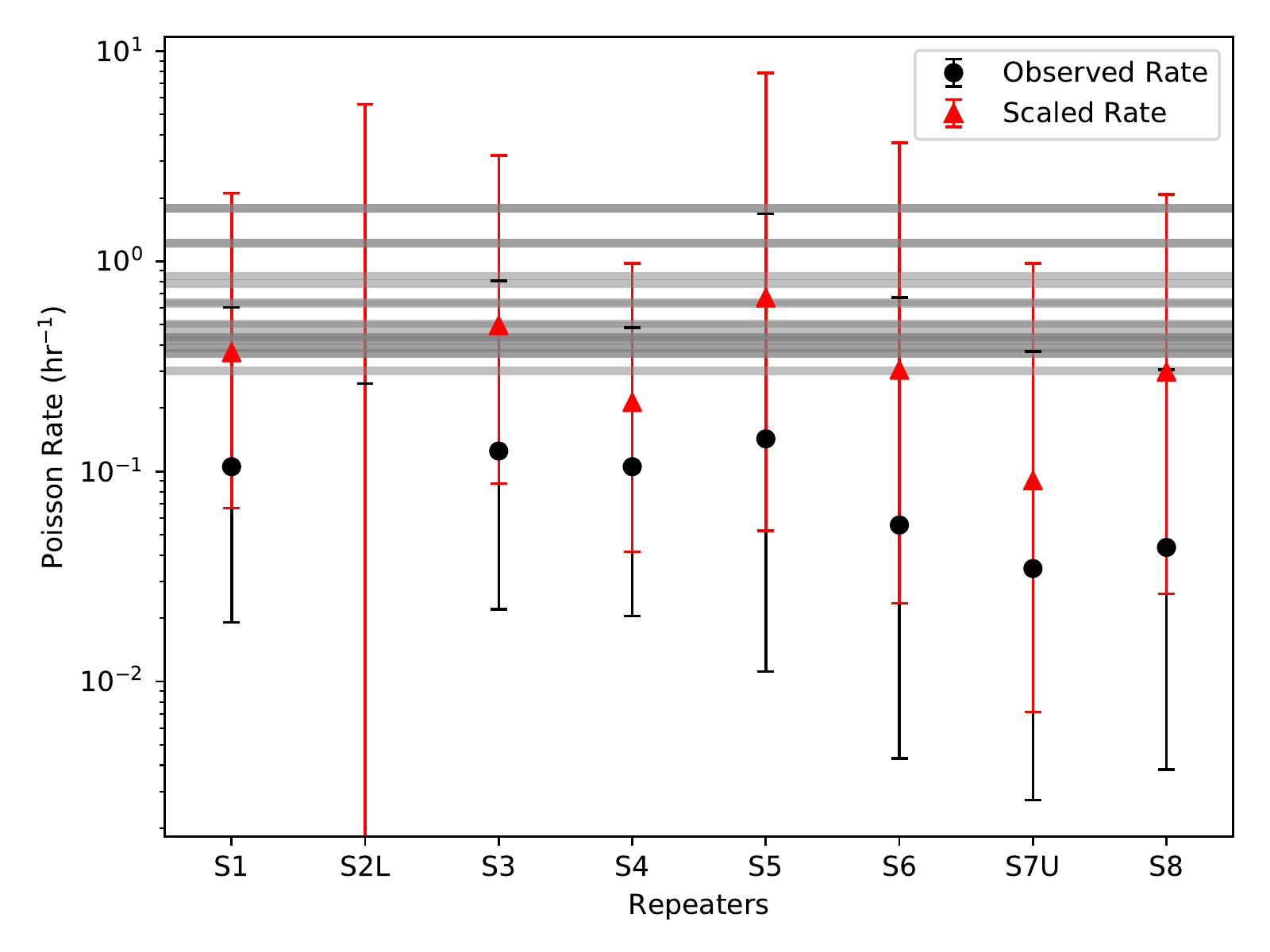}
	\end{center}
\figcaption{Repetition rates of CHIME/FRB repeaters. Observed rates (black circles) and sensitivity-scaled rates (red triangles) are shown with 68\% confidence interval Poisson error bars. Gray lines indicated the 68\% upper limits for the repetition rates of the 20 ASKAP-detected FRBs \citep{smb+18}, scaled to the sensitivity of CHIME/FRB. {\st For Source 2 and Source 7, only the rates in the lower and upper transits, where the fluence completeness thresholds 
{\vk are provided}
in Table~\ref{ta:repeaters}, are shown.}}
\label{fig:repetition_rates}
\end{figure}


\section{Conclusions}
\label{sec:conclusions}

We have reported on the discovery of eight new repeating FRB sources from CHIME/FRB. These include two sources with low DM, which we cannot exclude as being Galactic halo objects, and which are particularly promising for multi-wavelength follow-up once they are well localized.  One of them, Source 1, has RM = $-114.6 \pm 0.6$~rad~m$^{-2}$, much lower than what has been observed for FRB 121102. This, and the absence of a comparably luminous persistent radio source within the uncertainty regions of Sources 1 and 2, suggest that not all repeaters share the environmental properties of FRB 121102. Overall, the repeaters reported on here, together with 180814.J0422+73 \citep{abb+19b}, show no evidence of having DMs different from those of the so-far non-repeating CHIME/FRBs \citep{abb+19a}, but do show evidence of having larger burst widths.  This latter observation may suggest different emission mechanisms in repeating and non-repeating sources. We detect complex morphologies and downward-drifting sub-bursts in several -- but not all -- our events; our observations, along with those made of FRB 121102 \citep[e.g.,][]{hss+18}, strengthen the notion that such phenomenology is not necessarily observed in repeating sources, at least at $\sim$1-ms time resolution.


\bigskip

\acknowledgements

We are grateful for the warm reception and skillful help we have received
from the Dominion Radio Astrophysical Observatory, operated by the National Research Council
Canada. {\ef We thank Bradley W. Meyers for useful comments.} The CHIME/FRB Project is funded by a grant from the Canada Foundation for Innovation 2015
Innovation Fund (Project 33213), as well as by the Provinces of British Columbia and Que´bec, and by
the Dunlap Institute for Astronomy and Astrophysics at the University of Toronto. Additional support
was provided by the Canadian Institute for Advanced Research (CIFAR), McGill University and the
McGill Space Institute via the Trottier Family Foundation, and the University of British Columbia. The
Dunlap Institute is funded by an endowment established by the David Dunlap family and the University
of Toronto. Research at Perimeter Institute is supported by the Government of Canada through Industry
Canada and by the Province of Ontario through the Ministry of Research \& Innovation. The National
Radio Astronomy Observatory is a facility of the National Science Foundation operated under
cooperative agreement by Associated Universities, Inc.
We thank L. Piro for input to Figure 6.
M.B. is supported by a Fonds de recherche du Qu{\'ebec – Nature et technologies (FRQNT) Doctoral Research Award,  Physics Department Excellence Award, and a Mitacs Globalink Graduate Fellowship.
P.C. is supported by an FRQNT Doctoral Research Award and a Mitacs Globalink Graduate Fellowship.
M.D. is supported by the Watters Fellowship in Cosmology at UBC.
B.M.G. acknowledges the support of the Natural Sciences and Engineering Research Council of Canada (NSERC) through grant RGPIN-2015-05948, and of the Canada Research Chairs program.
A.S.H. is partly supported by the Dunlap Institute.
V.M.K. holds the Lorne Trottier Chair in Astrophysics \& Cosmology and a Canada Research Chair and receives support from an NSERC Discovery Grant and Herzberg Award, from an R. Howard Webster Foundation Fellowship from CIFAR, and from the FRQNT Centre de Recherche en Astrophysique du Quebec.
D.M. is a Banting Fellow.
M.M. is supported by a NSERC Canada Graduate Scholarship.
J.~M.-P. is supported by the MIT Kavli fellowship in Astrophysics and an FRQNT postdoctoral research scholarship.
Z.P. is supported by a Schulich Graduate Fellowship.
S.M.R. is a CIFAR Fellow and is supported by the NSF Physics Frontiers Center award 1430284.
P.S. is supported by a DRAO Covington Fellowship from the National Research Council Canada.
FRB work at UBC is supported by an NSERC Discovery Grant and by CIFAR. The baseband voltage system is funded in part by a John R. Evans Leaders Fund CFI award to I.H.S.

\bibliographystyle{aasjournal}

\bibliography{frbrefs}


\appendix
\section{Chance Coincidence Probability}
\label{app:chancecoincidence}

As the population of FRBs increases, so does the probability that two or more unrelated events happen to occur at the same sky location (within localization errors) and the same DM (within uncertainties). Here we calculate the chance coincidence probability $P_{cc}$ for the repeating FRB sources detected by CHIME/FRB. The probability of detecting multiple bursts within some phase space volume is,
$$
P_{cc}(\bar{V_0}, \Delta\bar{V}) = 1-p(0|\lambda(\bar{V_0})\Delta\bar{V}) - p(1|\lambda(\bar{V})\Delta\bar{V}), 
$$
where $\lambda(\bar{V})$ is the local background density of FRBs at a phase space location $\bar{V_0}$ and $\Delta\bar{V}$ is the phase space volume within which we would consider two bursts to be similar. $p(k|\lambda)$ is the Poisson probability mass function of getting $k$ events in an observation given an average rate $\lambda$. We choose to work in a phase space of declination $\delta$ and the excess DM (DM$_{ex}$), marginalizing over time of detection, RA, scattering time, and pulse width, because $\delta$ and DM$_{ex}$ are the strongest contributors to the variation in $\lambda$. The variation of sensitivity with zenith angle, the increasing transit time towards the North celestial pole, and the instrumental sensitivity variation with DM are automatically accounted for. Marginalizing over time of detection is necessary to account for the changes in system sensitivity, the on-off segments and the fact that repeaters can have a large range of repetition rates. 

The full sample of CHIME/FRB detected bursts will be discussed in upcoming papers. For the purposes of this calculation, we estimate $\lambda(\delta,{\rm DM}_{ex})$ by considering all the FRBs with detection S/N $>$ 9 with saved intensity data that have passed visual verification. We removed the multiple repeat bursts from FRB 180814.J0422+73, Sources 1 and 3, as well as carefully verifying the low excess DM events for contamination by pulses from known, ms-duration Galactic radio transients. We binned the number of FRBs within  $-10^\circ \leq \delta \leq 90^\circ$ in $2^\circ$ intervals and in $0\,\mathrm{pc\,cm^{-3}} \leq {\rm DM}_{ex} \leq 3500\,\mathrm{pc\,cm^{-3}}$ in $50\,\mathrm{pc\,cm^{-3}}$ intervals. Each histogram bin is divided by the sky area of the ring at constant declination and the DM bin to calculate the density of FRBs in units of $\mathrm{deg^{-2}\,pc^{-1}\,cm^{3}}$. We then smoothed the resulting density with a 2D Gaussian kernel with $\sigma_\delta=10^{\circ}$ and $\sigma_{{\rm DM}_{ex}} = 250\,\mathrm{pc\,cm^{-3}}$. Figure~\ref{fig:frb_density} (left) shows the resulting map of $\lambda(\delta,{\rm DM}_{ex})$.


\begin{figure}[t]
	\centering
\includegraphics[clip=true, trim=0cm 0cm 0cm 0cm, width=0.48\textwidth]{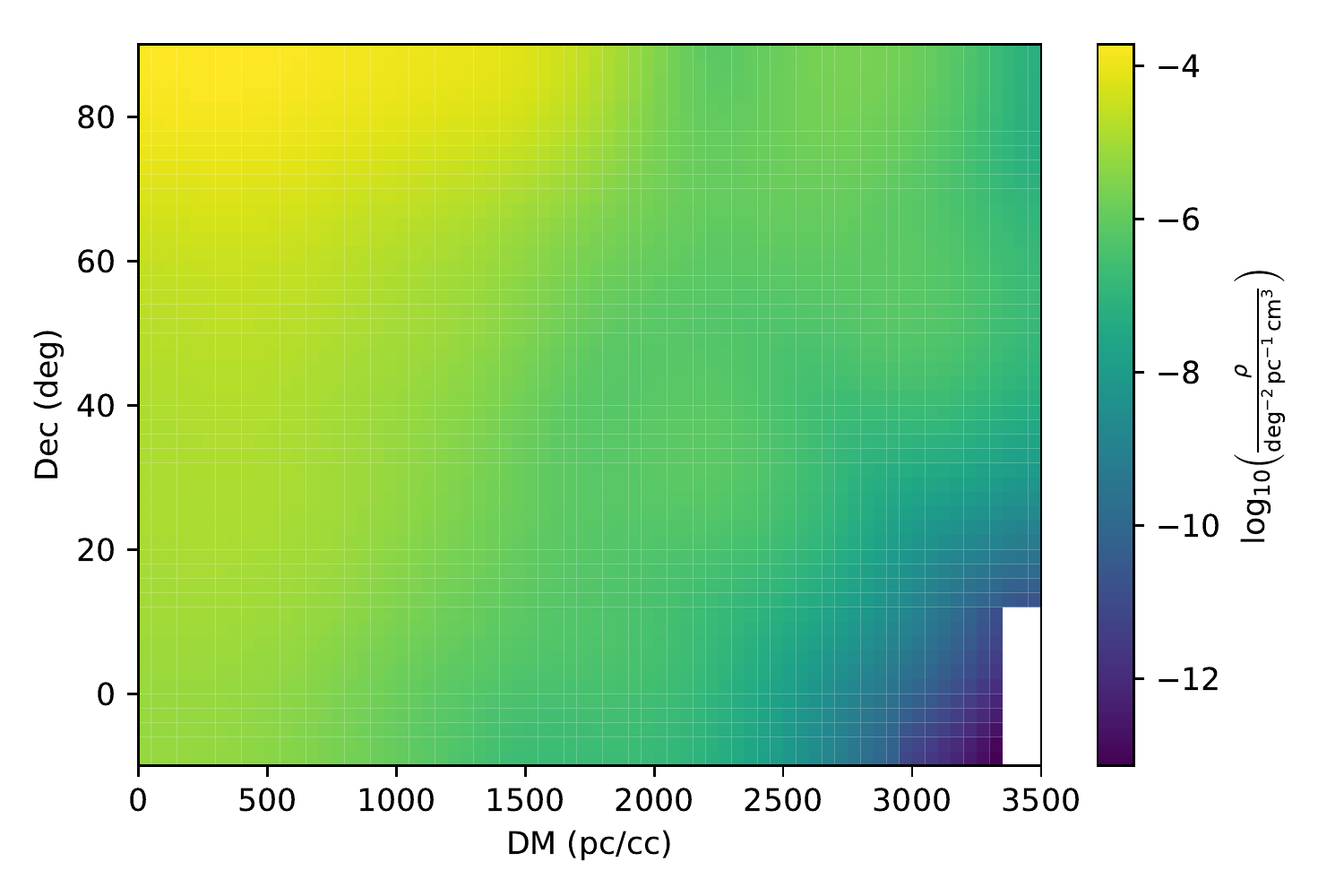}
\includegraphics[clip=true, trim=0cm 0cm 0cm 0cm, width=0.48\textwidth]{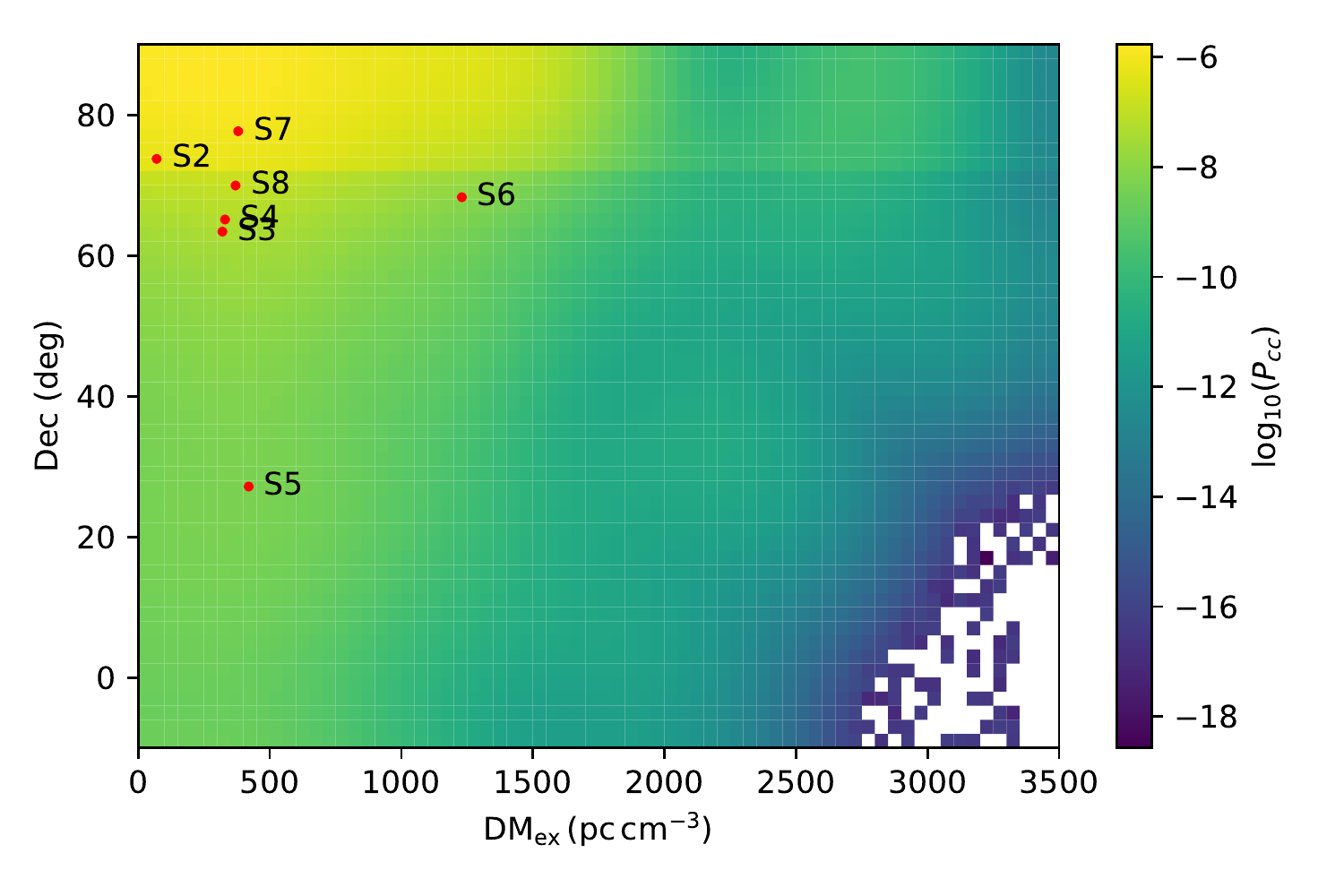}
\figcaption{Left: The areal density of FRBs in the declination and excess DM phase space. The color scale is logarithmic. Right: Chance coincidence probability for the repeaters detected by CHIME/FRB. The phase space positions of CHIME/FRB repeaters is overlaid.}
\label{fig:frb_density}
\end{figure}

For repeater identification, we calculate $P_{cc}$ if two or more bursts are within the same beam area and $\Delta {\rm DM}=10\,\mathrm{pc\,cm^{-3}}$. The beam area changes as a function of zenith angle. We calculate the area of the synthesized beam at each declination using the full width half maximum of the beam shape at 400\,MHz. Sources with $\delta>70^\circ$ are circumpolar in CHIME/FRB's field of view, with the transit at larger zenith angle having a larger beam area. To get a conservative estimate of $P_{cc}$, we use the larger beam areas at these declinations. Figure~\ref{fig:frb_density} (right) shows the resulting map of $P_{cc}$ with the phase space positions of the repeaters overlaid. For repeating sources with many more than two bursts detected, such as FRB 180814.J0422+73, Sources 1 and 3, $P_{cc}$ is far lower. 



\section{Exposure \& Sensitivity Estimation}
\label{app:exposure}

As discussed in detail in Section \ref{sec:exposure}, here we provide the timelines of the average daily exposures of the CHIME/FRB system to the sources listed in Table \ref{ta:repeaters}, for the interval from 2018 August 28 to 2019 February 25. The daily exposure for each source, shown in Figure \ref{fig:exposure}, corresponds to its transit across the FWHM region of the synthesized beams at 600 MHz with the average being computed over the positional uncertainty region. We include only the transits for which the CHIME/FRB system was fully operational {\PC --} i.e., the computing node designated for processing data for the synthesized beam corresponding to the transit was online and intensity data were being buffered to disk. 

We observe two daily transits from {\PC Sources 2 and 7} due to their circumpolar nature. For these sources, the daily exposure for the transit at lower elevation is evaluated separately since the sensitivity for the lower transit is significantly reduced as compared to that for the upper transit. Apart from the spatial variation in sensitivity due to the response of the primary beam, the sensitivity for each transit varies on a day-to-day basis due to changes in the RFI environment, software pipeline, and gain-calibration strategies. The corresponding daily variation in the RMS noise at the location of each source is reflected in the S/N detected with the CHIME/FRB system for Galactic pulsars transiting at elevation angles similar to that of the source. This variation in the RMS noise, along with the number of pulsars used for the daily measurement, is also plotted in Figure \ref{fig:exposure}. 

\begin{figure}
\gridline{\fig{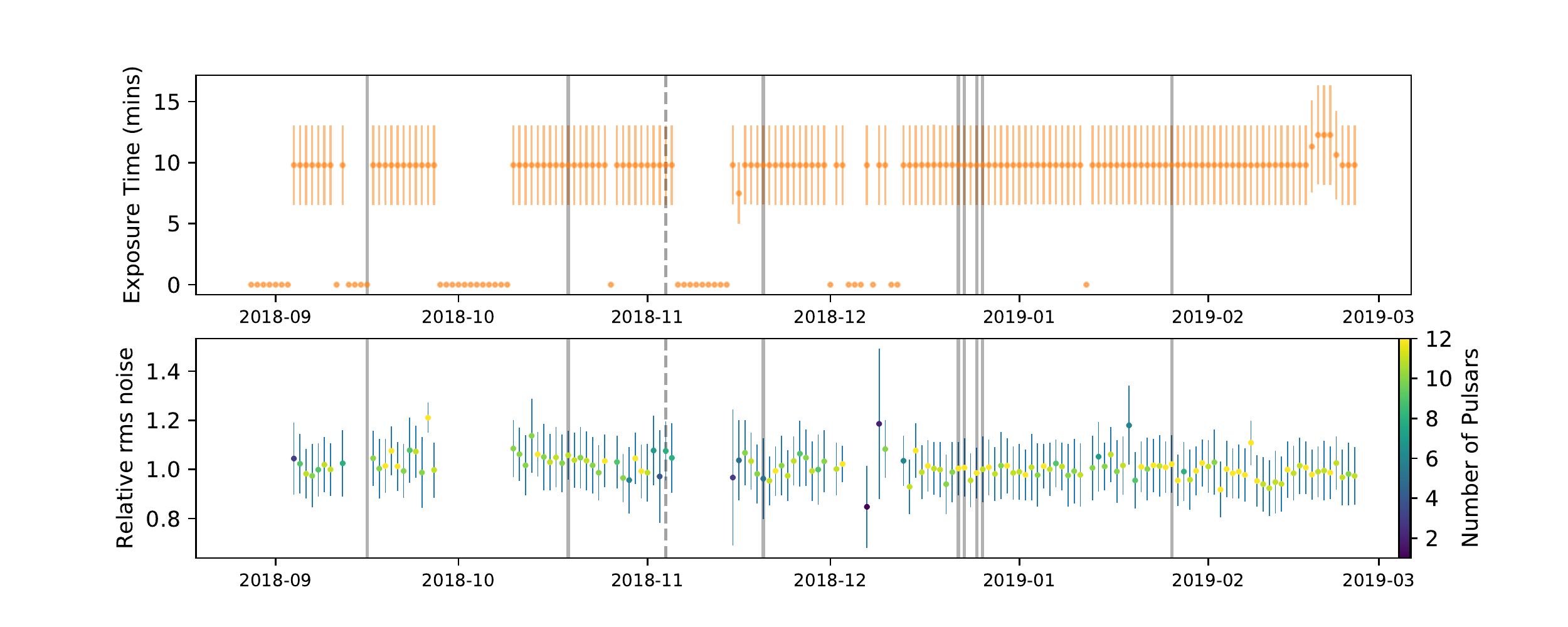}{0.85\textwidth}{(a) Source 1. Note that although a burst was detected on 2018 Sep 16, the plot indicates zero exposure for that day. This is because a surge in RFI in days prior to the detection interrupted recording of the system metrics, making it difficult to ascertain the exposure.}}
\gridline{\fig{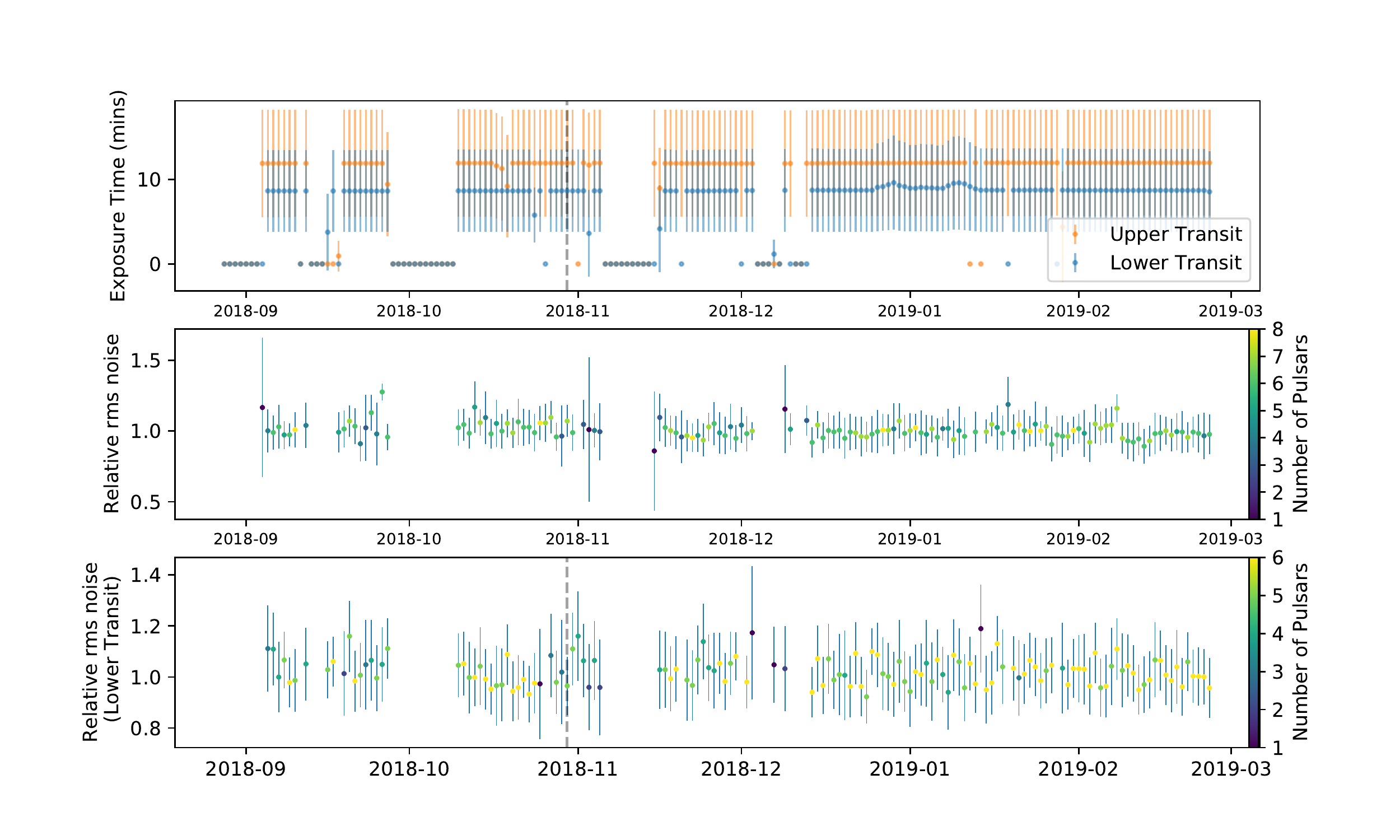}{0.85\textwidth}{(b) Source 2}} 
\caption{Timeline of CHIME/FRB's daily exposure to the new repeating FRB sources for upper and lower transits, if observable. Days on which a burst was detected are indicated by solid lines while dashed lines correspond to the detection of two bursts on the same day. The errors on the exposure are due to uncertainties in the source positions. The increase in exposure time from its typical value for some of the days is due to the occurrence of two transits in the same solar day caused by the length of a solar and a sidereal day being slightly different. The daily RMS noise at the position of each source is measured relative to the median for days having non-zero exposure to the source. This measurement is performed using pulsars detected by CHIME/FRB, the number of which is denoted by the marker colors.}
\label{fig:exposure}
\end{figure}
\addtocounter{figure}{-1}
\begin{figure}
\gridline{\fig{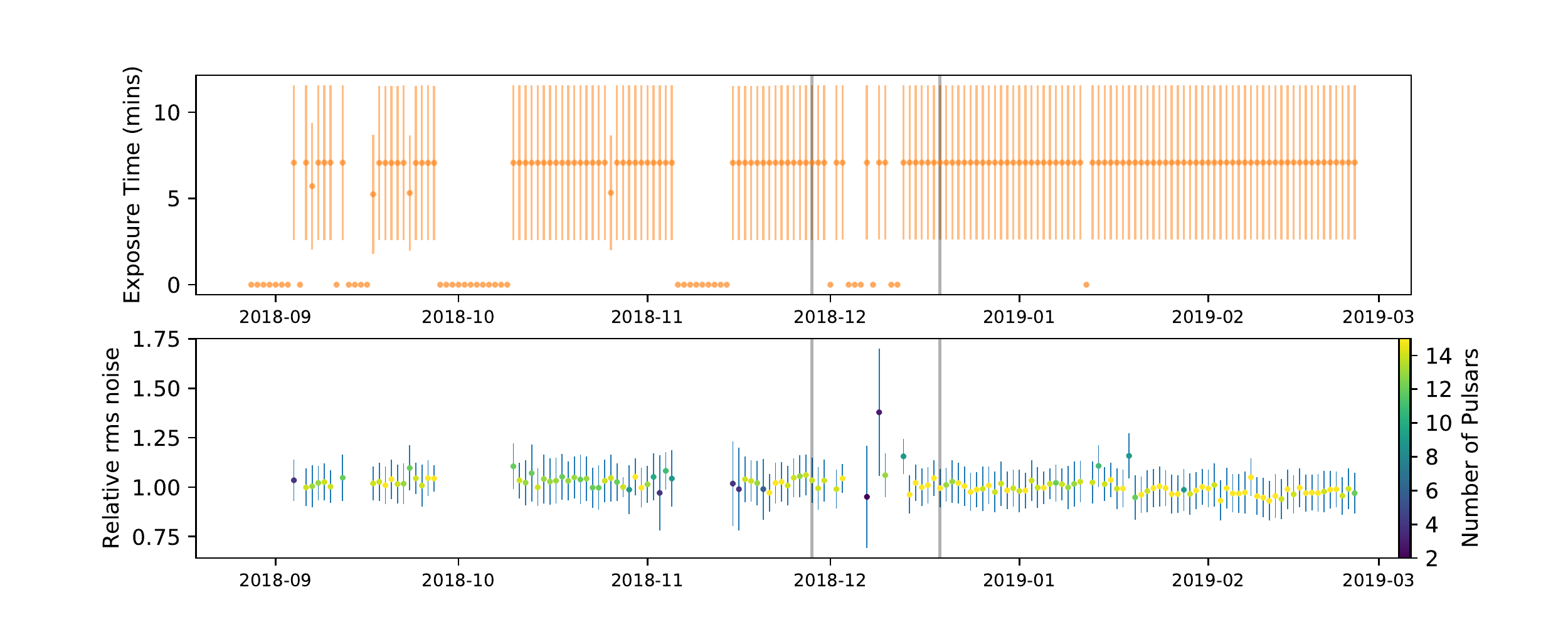}{0.85\textwidth}{(c) Source 3}}
\gridline{\fig{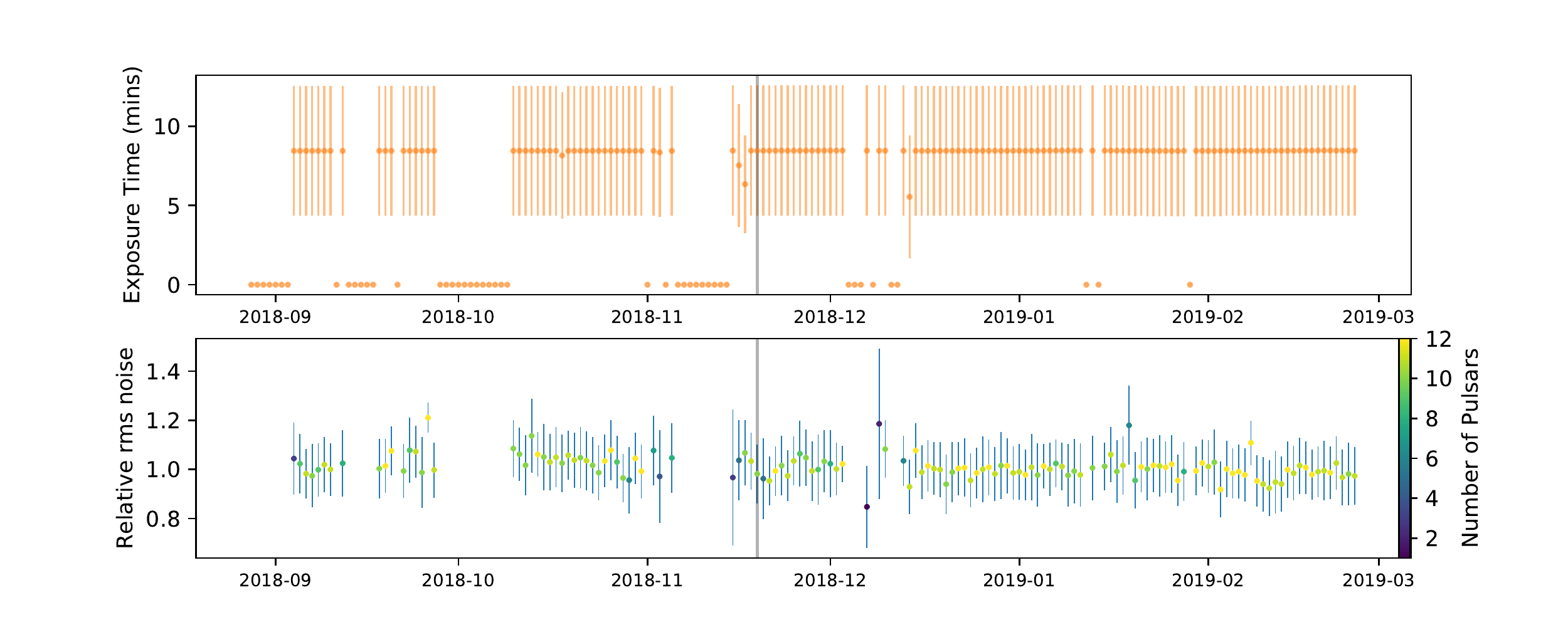}{0.85\textwidth}{(d) Source 4}}
\gridline{\fig{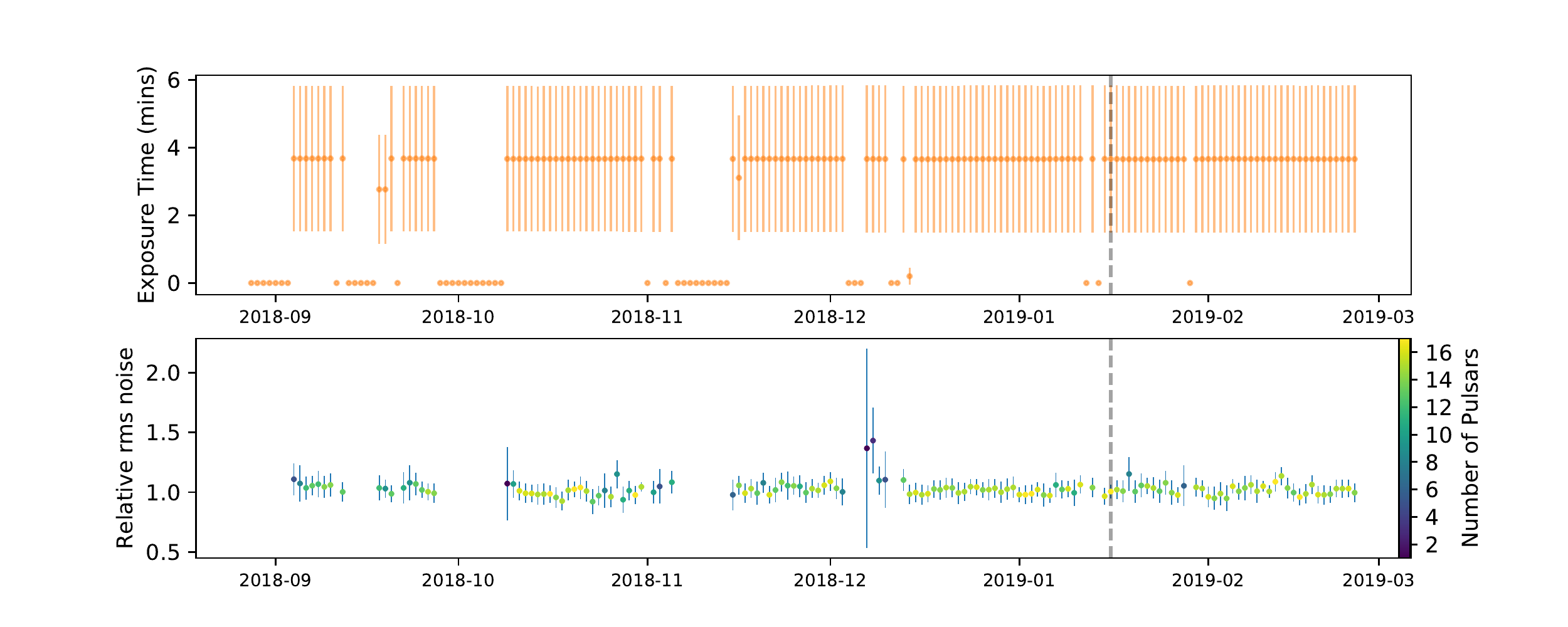}{0.85\textwidth}{(e) Source 5}}
\caption{Timeline of CHIME/FRB's daily exposure to the new repeating FRB sources. (cont.)}
\end{figure}
\addtocounter{figure}{-1}
\begin{figure}
\gridline{\fig{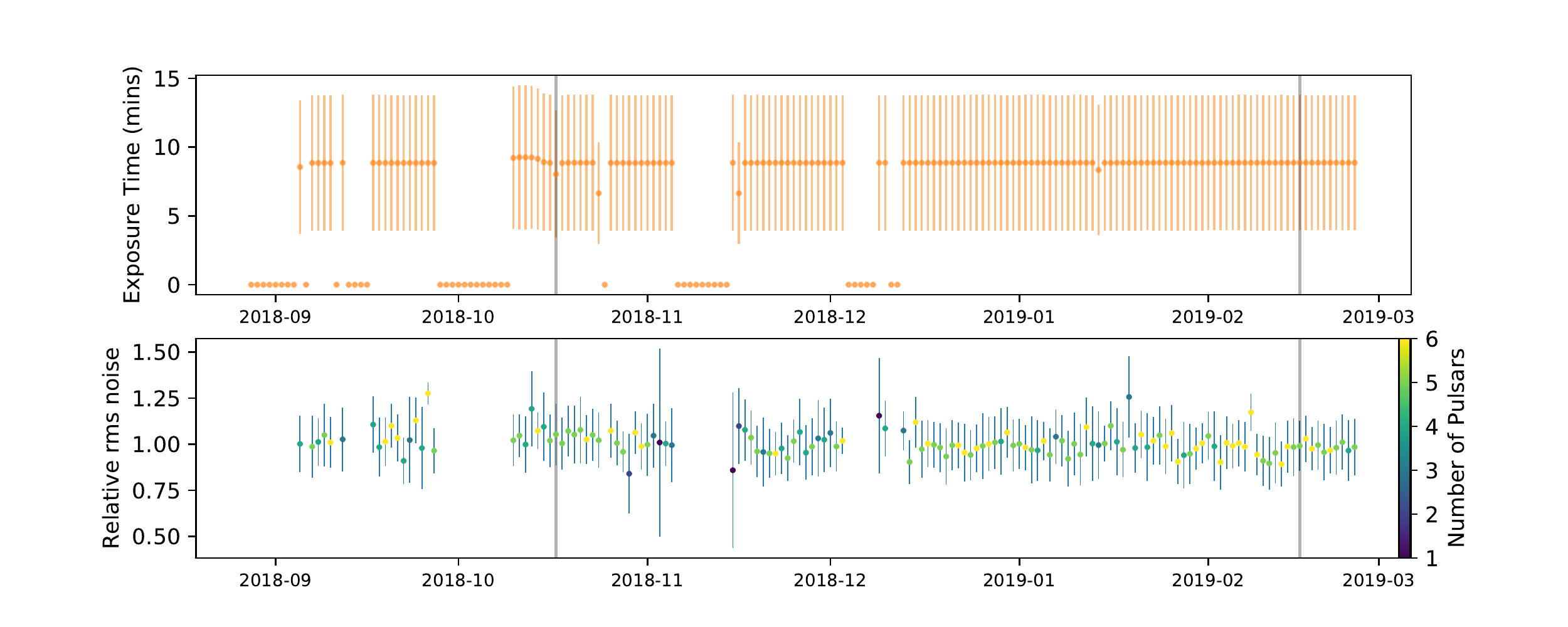}{0.85\textwidth}{(f) Source 6}}
\gridline{\fig{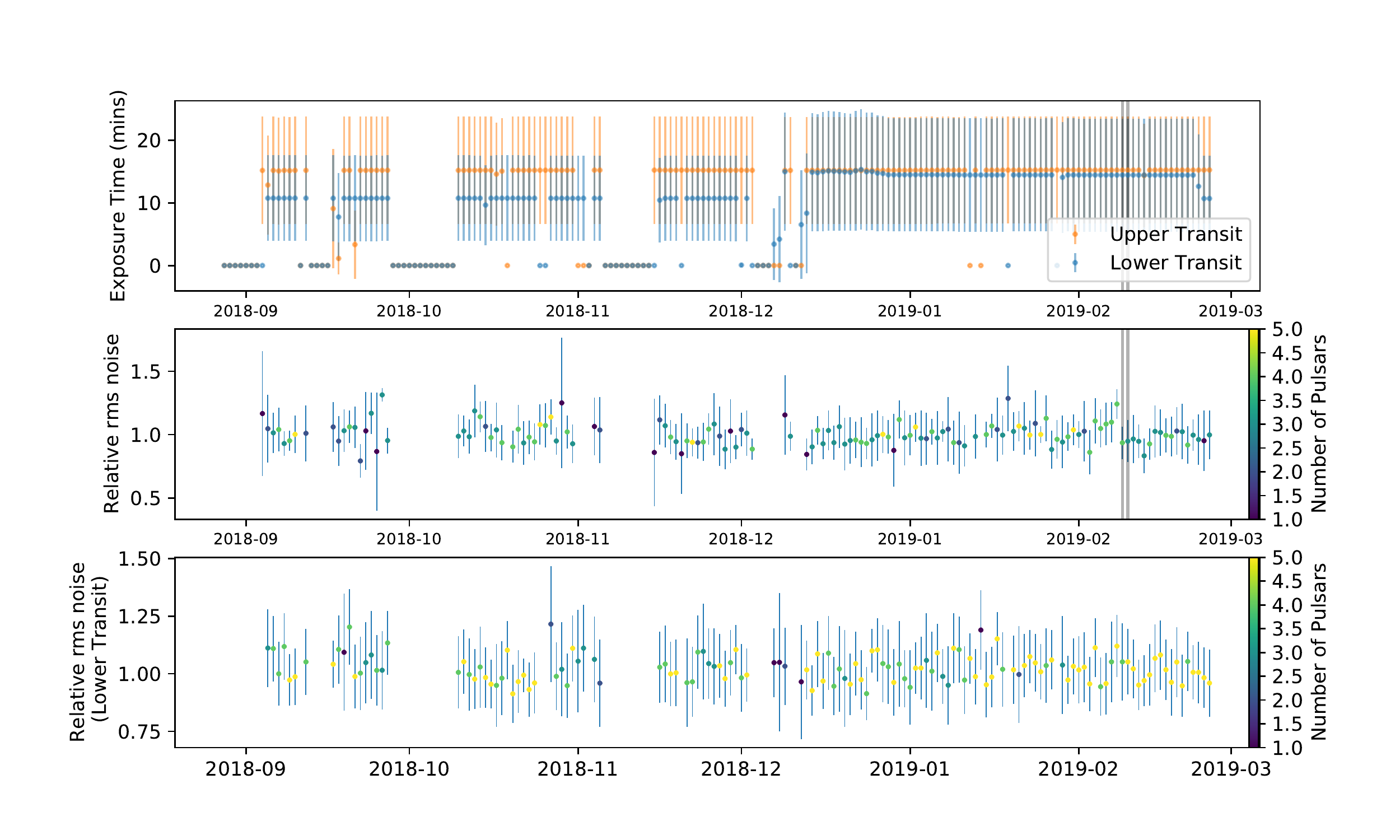}{0.85\textwidth}{(g) Source 7}}
\gridline{\fig{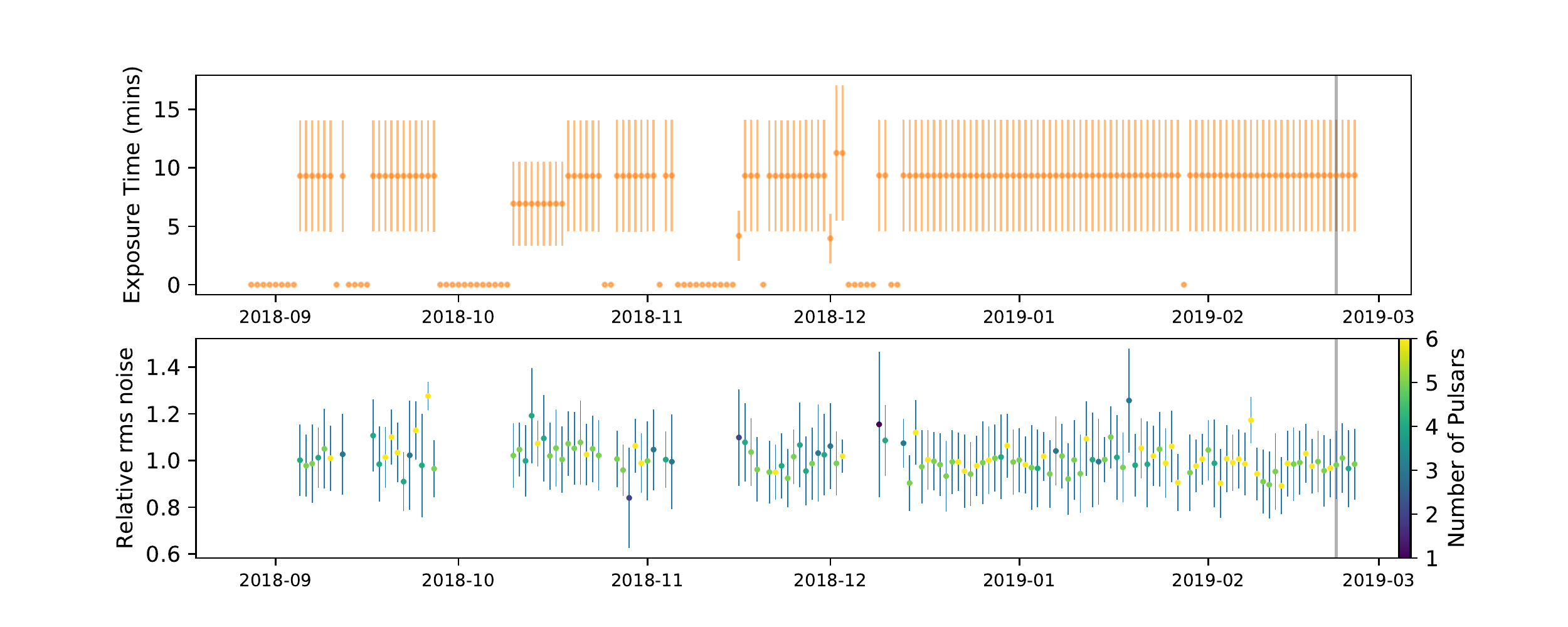}{0.85\textwidth}{(h) Source 8}}
\caption{Timeline of CHIME/FRB's daily exposure to the new repeating FRB sources. (cont.)}
\end{figure}
\end{document}